\documentclass[]{aa}

\usepackage{graphicx}
\usepackage{txfonts}
\usepackage{animate}
\begin{document}

   %\title{Structure and early evolution of embedded proto-clusters (I):\\ Fibers in the NGC1333}
   
   \title{Fibers in the NGC1333 proto-cluster  \thanks{Based on observations carried out under project number 169-11 with the IRAM 30m Telescope. IRAM is supported by INSU/CNRS (France), MPG (Germany) and IGN (Spain).}
   	\fnmsep\thanks{Based on observations with the 100-m telescope of the MPIfR (Max-Planck-Institut für Radioastronomie) at Effelsberg.}}
   
    %\title{Fibers in the NGC1333 proto-cluster}

  %\subtitle{Fibers in the NGC1333 proto-cluster}

   \author{A. Hacar
          \inst{1,2}
          \and
          M. Tafalla\inst{3}\and J. Alves\inst{1}%\and M. Lombardi\inst{1}
          }

   \institute{Institute for Astrophysics, University of Vienna,
              T\"urkenschanzstrasse 17, A-1180 Vienna, Austria
         \and
         	   Leiden Observatory, Leiden University, P.O. Box 9513, 2300-RA Leiden, The Netherlands \\
         	   \email{hacar@strw.leidenuniv.nl}
         \and
             Observatorio Astronomico Nacional (IGN), Alfonso XII, 3, E-28014, Madrid, Spain
             %\email{c.ptolemy@hipparch.uheaven.space}
             %\thanks{The university of heaven temporarily does not
             %        accept e-mails}
             }

   \date{Received ---, 2016; accepted ---, 2016}

\abstract{
Are the initial conditions for clustered star formation the same as for non-clustered star formation?  To investigate the initial gas properties in young proto-clusters we carried out a comprehensive and high-sensitivity study of the internal structure, density, temperature, and kinematics of the dense gas content of the NGC1333 region in Perseus, one of the nearest and best studied embedded clusters. The analysis of the gas velocities in the Position-Position-Velocity space reveals an intricate underlying gas organization both in space and velocity. We identified a total of 14 velocity-coherent, (tran-)sonic structures within NGC1333, with similar physical and kinematic properties than those quiescent, star-forming (aka fertile) fibers previously identified in low-mass star-forming clouds. These fibers are arranged in a complex spatial network, build-up the observed total column density, and contain the dense cores and protostars in this cloud.  Our results demonstrate that the presence of fibers is not restricted to low-mass clouds but can be extended to regions of increasing mass and complexity. We propose that the observational dichotomy between clustered and non-clustered star-forming regions might be naturally explained by the distinct spatial density of fertile fibers in these environments.
}


   \maketitle
%
%________________________________________________________________

\section{Introduction}

% Clusters
Clusters are the preferred sites for the formation of most of the stars in the Milky Way including massive stars \citep{LAD03}. The formation of stars in these compact systems is directly linked to the origin of the universal IMF, the stellar multiplicity and binary fraction, as well as the formation of planets \citep[e.g.,][]{BAT03}. Nevertheless, the origin of stars inside these compact environments remains under debate \citep[see][for recent reviews]{KRU14,LON14}.
 
% Observations
An extensive theoretical work demonstrates that the star formation properties within clusters
crucially depend on their initial gas conditions \citep[e.g.,][]{BON98,KRO01,DAL05,GOO06,MOE10,KRU12b}.
Two main factors limit the observational characterization of these objects. 
First, the initial gas properties within clusters (kinematics, density, temperature...) are rapidly altered by the direct impact of the stellar feedback, including outflows, winds, and radiation pressure \citep[e.g.,][]{DAL15}. 
As result, only the youngest, embedded proto-clusters are suitable for these type of studies.
Second, proto-clusters present a compact configuration of gas and stars, orders of magnitude denser than those of isolated star-forming regions. The interpretation of proto-cluster observations
is hampered by the intrinsically complex structure and kinematics involving gas along a wide range of physical conditions.

Of particular interest is the comparison between the star formation mechanisms in clusters and in isolation. Since the first millimeter observations of molecular clouds, it is well established that individual stars are originated within dense cores \citep{MYE83}.
As highlighted by recent Herschel continuum observations, most of the young stars and cores in low-mass regions are embedded in filaments of gas and dust dominating the cloud structure \citep{AND10,AND14}.
Molecular line observations reveal that some of the most prominent Herschel filaments are actually collections or bundles of sonic fibers \citep{HAC13,TAF15}. As demonstrated by the analysis of their internal gas kinematics, these quiescent fibers set the initial conditions for the gravitational collapse of cores within these isolated star-forming clouds \citep{HAC11,ARZ13,HAC16a}. Interestingly, high-resolution observations in clusters suggest the presence of an underlying gas substructure in regions of increasing complexity \citep[e.g.,][]{AND07,KIR13,FER14,HEN16}. While cores are routinely surveyed within these environments \citep[e.g.,][]{MOT98}, the identification of a pre-existing fiber-like substructure in clusters is, however,  subject of a strong controversy \citep[e.g.,][]{FRI16}.

% NGC1333
In this paper, we aim to investigate the internal gas substructure in the \object{NGC1333} proto-cluster.
After $\rho$-Oph, NGC1333 is the second nearest young proto-cluster \citep[D=238$\pm$18~pc][]{HIR08}. 
Ought to its proximity and northern declination ($\delta=+31.5\degr$), the NGC1333 region 
has been investigated along the entire electromagnetic spectrum in the last decades, becoming one of the best characterized star-forming regions in the solar neighbourhood \citep[see][for a review]{WALA08}.
Its stellar population has been extensively surveyed in the optical \citep{RAC68}, IR \citep{STR74,JEN87,LAD96,JOR06,GUT08,FOS15}, FIR \citep{ENO09,SAD14}, radio continuum \citep{TOB16}, and X-rays \citep{PRE97,GET02}.
The prominent star formation of this region is recognized by its intense outflow activity \citep{BAL96}, widely characterized in the past using (sub-)millimeter and IR observations \citep{KNE00,HAT07,HAT09,ARC10,PLU13,DIO16}.
Similarly, the gas content in NGC1333 has been investigated at large scales combining IR extinction \citep{LOM10}, as well as FIR \citep{SAD14,ZAR16}, millimeter continuum \citep{SAN01,LEF98,HAT05,KIR06,ENO07}, and line observations \citep{WAR96,RID06,CUR11}. Dedicated surveys have investigated both the core population \citep{JOH10,ROS08} and dense gas properties of this cloud in high detail \citep{WAL06,WAL07}.
Here, we present the analysis of a new set of high-sensitivity, large-scale observations both N$_2$H$^+$ and NH$_3$ density selective tracers along the entire NGC1333 region. Combined with previous IR surveys and archival data, we aim to fully characterize the dense gas properties (distribution, mass, density, temperature, and kinematics) in this proto-cluster. 

The paper is organized as follows. First, and  in Sect.~\ref{observations}, we present the millimeter line and FIR continuum observations used in this work. In Sects. \ref{sec:NGC1333} and \ref{sec:densegas}, we investigate the mass content, mass distribution, plus the gas and dust thermal properties of the NGC1333 cluster. Similarly, we also explore the connection between the position and distribution of dense gas with the distribution of the newly formed stars and derive efficiencies and timescales for its evolution. Section~\ref{sec:kinematics} is devoted to the detailed analysis of the gas velocity field and the detection of fibers in NGC1333.
Finally, and in Sect.~\ref{sec:discussion}, we compare these new results with the structure and properties of the gas found for isolated star forming regions and discuss their implications for our current description of the star formation process in molecular clouds.

\section{Observations and data reduction}\label{observations}

\begin{figure*}
	\centering
	\includegraphics[width=15cm]{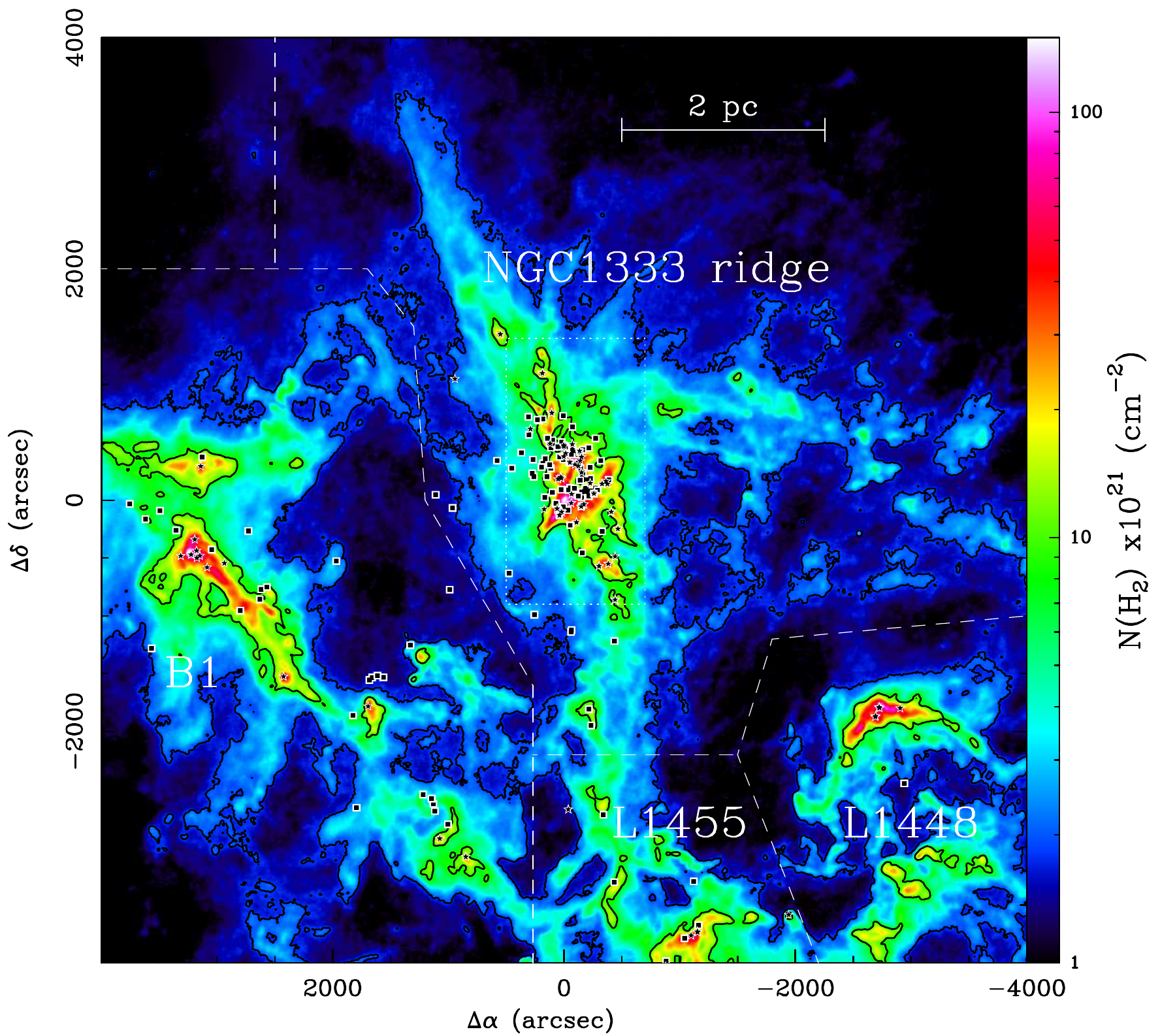}
	\caption{Large scale view of the NGC1333 ridge in Perseus. The image represents the total column density map derived from our Herschel-Planck maps (in color scale), including two contours of equivalent column densities of A$_V=$~2$^{\mathrm{mag}}$ and 10$^{\mathrm{mag}}$, respectively. Offsets are referred to the map centre with coordinates $(\alpha,\delta)_{J2000} = (03^h29^m08^s.9,+31\degr15'12'')$ in radio projection. 
		The 4 subregions around NGC1333 and their boundaries are labelled in the plot. 
		The different symbols correspond to the Class~0/I (stars) and Flat/ClassII/III objects (squares) identified by different Spitzer surveys \citep{GUT08,EVA09}.
		The dotted line encloses the NGC1333 clump studied in this work.}
	\label{NGC1333_ridge}
\end{figure*}

% IRAM30m
The bulk of the data used in this work corresponds to different observations carried out with the IRAM~30m telescope of the central clump of NGC1333 during December 2011 and March 2012. We observed this region at the frequency of the N$_2$H$^+$ (JF$_1$F=123--012) line \citep[93173.764 MHz,][]{PAG09} using the EMIR receiver. We connected this front-end to the VESPA autocorrelator configured to provide a spectral resolution of 20~kHz, equivalent to 0.06~km~s$^{-1}$ at the frequency of the N$_2$H$^+$ (1-0) line.
The observations consist of a large mosaic of 35 submaps with sizes between 100$\times$100 and  200$\times$200 arcsec$^2$ each, covering a total area of $\sim$~340 arcmin$^2$, centred at the position $(\alpha,\delta)_{J2000} = (03^h29^m08^s.9,+31\degr15'12'')$  (i.e., Core~73 of \citet{ROS08}). Each submap was observed multiple times and in orthogonal directions in On-the-fly (OTF) and frequency-switching (FSw) mode, with a scan velocity of v$_{\mathrm{scan}}=$~5~arcsec~s$^{-1}$, a dump time of t$_{\mathrm{dump}}=$~1~s, a row spacing of L$_{\mathrm{rows}}=$~5~arcsec, and a frequency throw of $\nu_{\mathrm{throw}}=\pm$4.5 MHz. Additionally, and also in FSw mode, we obtained deep integrations in 70 independent positions along the cloud. Both sky calibrations and pointing corrections were obtained every $\sim$~10-15 min and $\sim$~1.5~hours, respectively. Conversions between antenna and main beam temperatures assumed a standard telescope main beam efficiency of $\eta_{mb}=0.84$\footnote{Value 
obtained interpolating facility-provided forward and beam efficiencies at the N$_2$H$^+$ (1--0) frequency. See also http://www.iram.es/IRAMES/mainWiki/Iram30mEfficiencies .}. 
The stability of the receiver and the cross-calibration between different sessions were regularly checked using deep observations of the central position and were found to be better than 15\%. 

% EFF100m
During November 2011, we also targeted the NGC1333 proto-cluster with the Effelsberg 100m radiotelescope. We  mapped this cloud in both NH$_3$ (1,1) and (2,2) inversion lines simultaneously at 23694.495 and 23722.633 MHz \citep{KUK67},respectively, using the P1.3mm front-end and FFT facility spectrometer set to a spectral resolution of 6~kHz or 0.08~km~s$^{-1}$.
With a final coverage and an observational strategy analogous to the N$_2$H$^+$ (1--0) maps, a large mosaic was obtained combining different submaps of 200$\times$200 arcsec$^2$. Each tile was observed combining orthogonal directions and carried out in OTF and FSw modes with parameters v$_{\mathrm{scan}}=$~5~arcsec~s$^{-1}$, t$_{\mathrm{dump}}=$~2~s,  L$_{\mathrm{rows}}=$~20~arcsec, and $\nu_{\mathrm{throw}}=\pm$2 MHz.
In addition to these maps, high-quality spectra were obtained in 8 position along this region.
Similar to the IRAM30m observations, pointing and focus corrections were carried out every 1.5-2~hours.
Conversely, these data were calibrated offline and in units of main beam temperature. For that,  
all the spectra were first converted into antenna temperature units assuming typical T$_{\mathrm{cal}}$ conversion factors,
then corrected by atmospheric attenuation and gain-elevation
applying standard calibration procedures\footnote{See Kraus (2007) at https://eff100mwiki.mpifr-bonn.mpg.de}, and finally corrected by a main beam efficiency of $\eta_{mb}=0.59$.
To test the quality of our data and the reduction process, different measurements of NH$_3$~(1,1) line at the central position of the L1551 core were obtained during our observations.
% ($(\alpha,\delta)_{J2000} = (04^h31^m34^s.08,+18º08'05'')$).
Relative and absolute measurements of the main beam temperature at the central position of this core were compared to the calibrated results obtained by \citet{MEN85} finding differences less than 30\% in all cases.

% reduction and convolution
A total of $\sim$~380\,000 individual calibrated spectra for the N$_2$H$^+$ (1--0) and the NH$_3$ (1,1) and (2,2) lines were produced in our OTF IRAM~30m and Effelsberg~100m observations at the native telescope FWHM of $\sim$~26 and 42~arcsec, respectively. The data were reduced using GILDAS/CLASS software\footnote{http://www.iram.fr/IRAMFR/GILDAS}. First, the different datasets were independently combined, convolved, and Nyquist resampled with a final resolution of 30~arcsec for the N$_2$H$^+$ and 60~arcsec in the case of the NH$_3$ lines. After that, a baseline correction was applied to all the resulting spectra after subtracting a third-order polynomial. 
The final maps resulted in a total of 5525 spectra of N$_2$H$^+$ (1--0) and 1006 spectra for each of the NH$_3$ (1,1) and (2,2) lines, all with typical rms values of $\sim$~0.15 K \footnote{All the molecular data used in this paper are available via CDS}.

% Herschel-Planck
In addition to our molecular observations, we made use of the archival high-resolution and high-dynamical range total column density (N(H$_2$)) and dust effective temperature (T$_{\mathrm{eff}}$) maps presented by \citet{ZAR16}. These maps were obtained from the combination of Gould-Belt Herschel Key-project data \citep{AND05} calibrated against both all-sky Planck galactic maps \citep{PLA11b} and the 2MASS-NICER extinction maps \citep{LOM10} of the Perseus molecular cloud.
The values for the 350$\mu$m dust opacities ($\tau_{350\mu m}$) derived by Zari et al were transformed into their corresponding K-band extinction (A$_K$) following the conversion factors defined by these authors. Every A$_K$ measurement was converted first into its corresponding visual extinction value (A$_V$) and then into its total H$_2$ column density  adopting standard reddening law \citep[$\mathrm{A}_K/\mathrm{A}_V=0.112$, ][]{RIE85} and dust-to-gas conversion factors \citep[$\mathrm{N}(\mathrm{H}_2)/\mathrm{A}_V=0.93\times10^{21}$ cm$^{-2}$ mag$^{-1}$, ][]{BOH78}.  As a result, these data provide us with fully calibrated N(H$_2$) and T$_{\mathrm{eff}}$ maps of the NGC1333 region with a final resolution of 36~arcsec.
Errors in both N(H$_2$) and T$_{eff}$ measurements were derived by \citet{ZAR16} and are estimated on $\lesssim$~10\%.

%%%%%%%%%%%%%%%%%%%%%%%%%%%%
%% Large scale
%%%%%%%%%%%%%%%%%%%%%%%%%%%%
\section{Dust continuum emission: Large scale properties}\label{sec:NGC1333}

\begin{figure*}
	\centering
	\includegraphics[width=\textwidth]{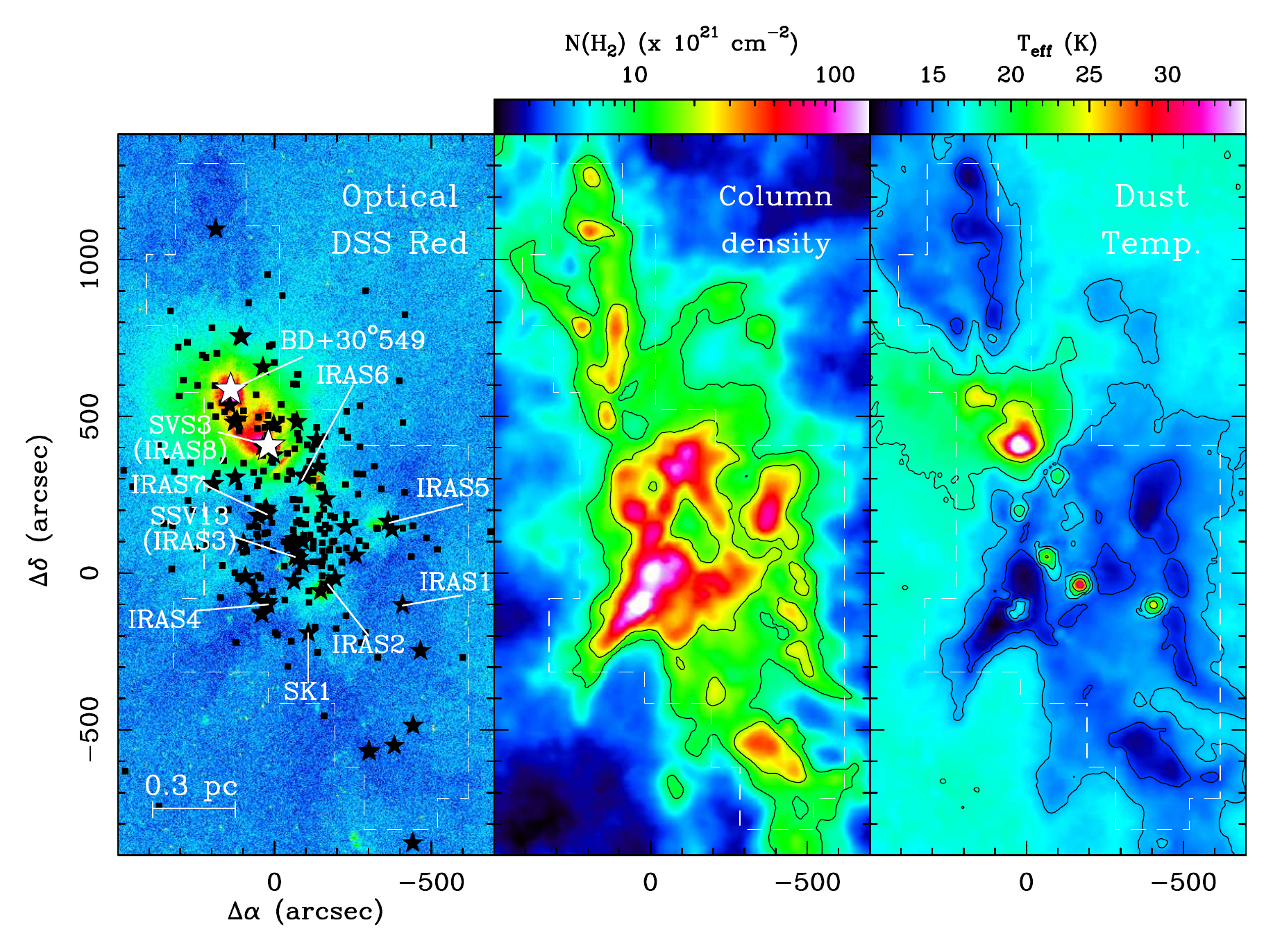}
	\caption{Optical DSS Red image (left) of the NGC1333 clump compared to the Herschel-Planck total column density (centre) and dust effective temperature (right) maps \citep{ZAR16}.
		Offsets are referred to the map centre with coordinates $(\alpha,\delta)_{J2000} = (03^h29^m08^s.9,+31\degr15'12'')$ in radio projection.
		The position of the two B-type stars (solid white stars), as well as all the Class 0/I protostars (solid black stars) and Class Flat/II/III objects (solid black squares) identified by \cite{REB15}, are indicated in the first subpanel. The black contours correspond to extinction values of A$_V=$~10 and 20$^{\mathrm{mag}}$ and dust effective temperatures of T$_{\mathrm{eff}}=$~14, 16, and 18~K in their respective maps. For comparison, the area covered by our IRAM30m observations is indicated by a dashed line in all plots (see also Fig.~\ref{NGC1333_densetracers}).}
	\label{NGC1333_largescale}
\end{figure*}

\subsection{The NGC1333 ridge}\label{subsec:NGC1333ridge}

% Large scale description (ridge)
In Fig.~\ref{NGC1333_ridge}, we present a large scale view of the NGC1333 region. 
In good agreement to the previous extinction \citep[e.g.,][]{LOM10} and continuum maps \citep{HAT05,ENO06,SAD13}, this figure shows the distribution of the total gas column density around NGC1333 derived from our Herschel-Planck maps. The NGC1333 proto-cluster lies at the centre of a $\sim$~2~pc length and elongated dense clump found at A$_V\ge$~10$^{mag}$.
This clump belongs to a diffuse gas ridge identified at A$_V\ge$~2$^{mag}$, running approximately north-south along $\sim$~6~pc in the middle of this image. NGC1333 is surrounded by another three well-known star-forming regions, namely B1, L1448, and L1455, clearly separated at low column densities.

% SF and YSOs distribution
Figure~\ref{NGC1333_ridge} also shows the distribution of YSOs in the surroundings of the NGC1333 ridge. A total of $\sim$~180 Spitzer objects are identified within the boundaries of this map \citep{JOR06,EVA09}. Counted together, NGC1333, B1, L1448, and L1455 contain about half of the total YSO population and $\sim$~85\% of the protostars identified in the entire Perseus cloud. Only NGC1333 contains $\sim$~70\% of all these objects, most of them within a radius of 1~pc around its central clump. Indeed, NGC1333 is the second most active star-forming region in Perseus after IC~348. NGC1333 contains, nevertheless, two times more Class~0/I objects than IC~348, denoting its current star formation activity  \citep{JOR08}. Combining the information of dedicated surveys at X-rays, optical, and radio wavelengths, \citet{REB15} identified an total of 277 potential cluster members within the NGC1333 region. 
The relative large number of Class II/III objects suggests that the star formation process in NGC1333 has been active for the last $\sim$~2-3~Myrs. 

% Mass of the cloud: comparisons
The current star-formation activity in NGC1333 is likely related to its large content of gas at high column densities.
Within the  A$_V=$~2$^{mag}$ contour defined in Fig.~\ref{NGC1333_ridge}, we estimate a total mass of $\sim$~1700 M$_\sun$ for the entire NGC1333 ridge, calculated adding the contribution of all the pixels within this contour assuming a mean molecular weight of $\mu=$~2.33. 
Similarly, we also estimate a total of $\sim$~580~M$_\sun$ for the gas within NGC1333 at A$_V\ge$~10$^{mag}$, most of them concentrated in its central clump.
Compared to its neighbours, the total mass of the NGC1333 ridge presents about twice the total mass found in L1455 and L1448 clouds. Both NGC1333 and B1 show roughly similar masses at low (A$_V\ge$~2$^{mag}$) and intermediate (A$_V\ge$~10$^{mag}$) column densities, while their distribution significantly differ only at relatively high extinctions (A$_V>$~30$^{mag}$). These last differences might explain their distinct star-formation histories.

\subsection{Mass distribution within the NGC1333 clump}\label{subsec:HP_mass}

% Figure 2, optical image and YSOs population
Figure~\ref{NGC1333_largescale} shows a close-up view of the NGC1333 central clump.
Superposed to the optical DSS-Red image (left), we present the distribution of the 9 IRAS sources \citep{JEN87} and the 277 YSOs candidates \citep{REB15} identified within this region.
As seen in this figure, the stellar population of the NGC1333 cluster is dominated by the presence of two late-type B stars, BD+30$\degr$549 (IRAS9; B8V) and SVS3 (IRAS8; B5e) \citep{STR74,CER90}. 
Observed in scattered light in the optical, both stars are exciting an intense reflection nebula \citep{VdB66,RAC68}. 
As pointed out by different studies \citep{LAD96,GUT08}, the stellar population in NGC1333 is segregated in two groups, typically referred to as North and South subclusters, respectively (see Sect.~\ref{subsec:dense_proto}). The North subcluster, centred at the position of (x,y)$\sim$~(50,400) arcsec in our maps, is located at the surroundings of SVS3 and is dominated by a large population of Class II/III objects. Conversely, the South subcluster is centred around (x,y)$\sim$~(-100,100) arcsec and contains most of the young sources found in this region, like IRAS~2, IRAS~4, and SSV13 (IRAS~3). 

% Figure 2, column density map
Figure~\ref{NGC1333_largescale} also illustrates the clumpy distribution of gas inside NGC1333 revealed in the continuum \citep{LEF98,SAN01,HAT05,ENO06}.
Most of the high column density material is located in the middle and southern parts of NGC1333, forming an intricate network of cores and filamentary structures (see also Sect.~\ref{subsec:kinematics:fibers}). 
The most prominent condensations of gas are found toward to the position of the youngest IRAS sources. Several of these regions, like in the case of IRAS~3 or IRAS~4, are found at A$_V>100^{mag}$.
According to our analysis of the kinematics (Sect.~\ref{sec:kinematics}), these extraordinary column densities are produced by the superposition of multiple dense gas components along the line-of-sight.

In addition to the previous ground-based continuum observations, the improved sensitivity of our Herchel-Planck maps allow us to describe the gas distribution at low extinctions. As seen in Fig.~\ref{NGC1333_largescale}, the gas column density in NGC1333 rapidly decreases outside its central clump. Most of the YSOs identified in these region, are enclosed and connected by a contour of A$_V\sim~10^{mag}$. The gas distribution traced by these contours also defines a ring-like region centred at position of SVS3. The coincidence of the rim of this structure with the extension of the optical reflection nebula suggests that they are likely tracing the edges of the cavity eroded by this still partially embedded star.

 \begin{figure}
   \centering
   \includegraphics[width=\linewidth]{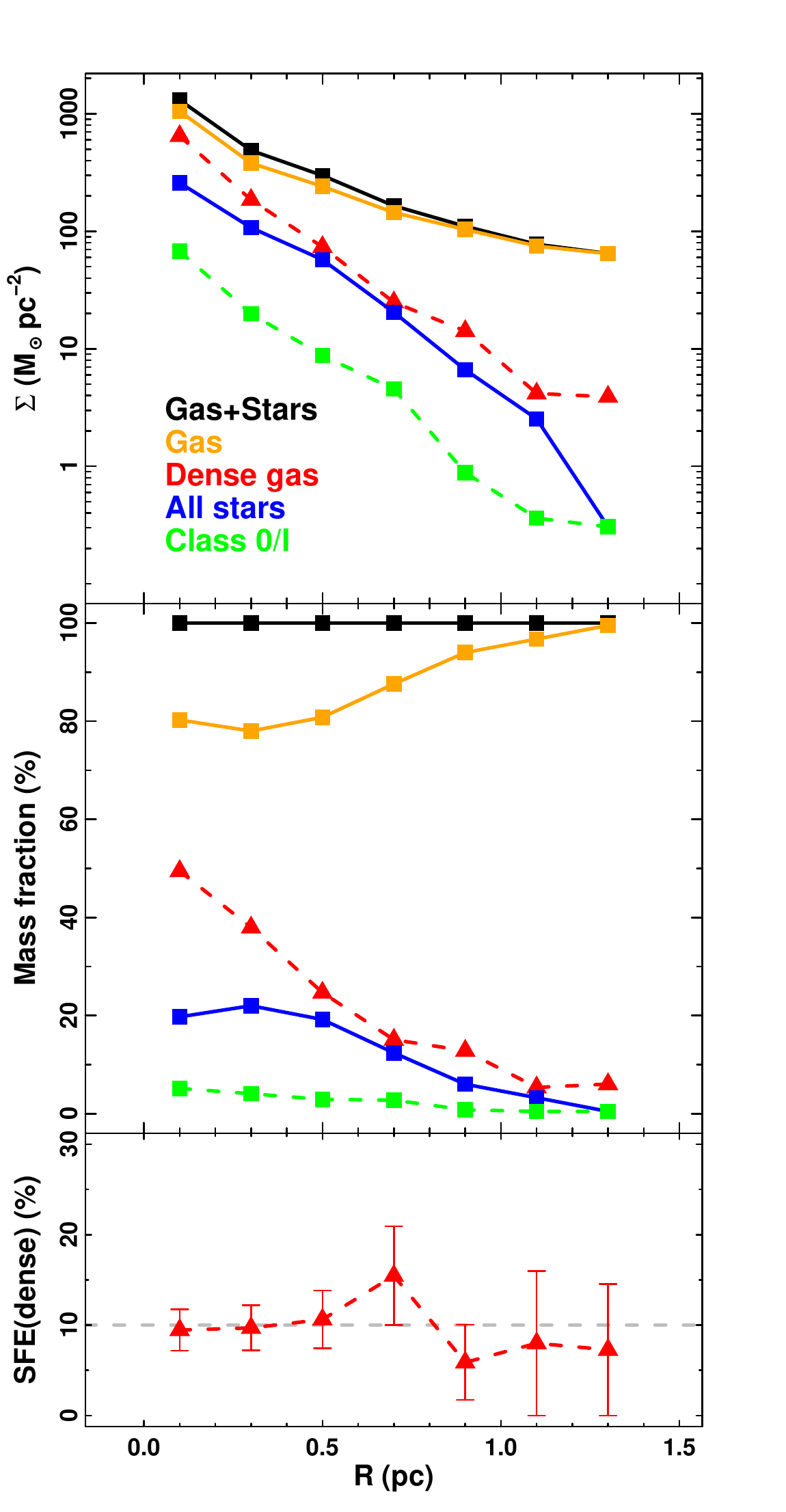}
      \caption{Mass surface density ($\Sigma$; upper panel), relative mass fraction (mid panel), and Star formation Efficiency for dense gas (lower panel) as a function of the impact parameter R with respect to the centre of the NGC1333 cluster (see text). The different physical components of gas (orange), dense gas (red), Stars (Classes 0-III; blue), Class 0/I protostars (green), and total mass (i.e. Stars+Gas; black) are color coded in the corresponding plots.}
             \label{NGC1333_massdistrib}
   \end{figure}

% Surface density and mass distribution
The detailed information provided by the different Spitzer surveys combined with our continuum maps offers an opportunity to study the radial distribution of the mass in both stellar and gas components inside the NGC1333 proto-cluster.
We estimate the mass surface density at a given impact parameter $R_i$ from its centre ($\Sigma\ (R_i)$) from the contribution of all the mass elements within an annulus $R_i \in [R_i-\delta/2, R_i+\delta/2) $ assuming circular symmetry:
\begin{equation}
  \Sigma (R_i) = \frac{1}{\pi \left[(R_i+\delta/2)^2-(R_i-\delta/2)^2\right] }\sum\limits m (R_i) \label{eq:surfden}
\end{equation}
For practical reasons, we identify the centre of the NGC1333 cluster with the centre of our molecular maps. 
Using Eq.~\ref{eq:surfden}, we have calculated the mass surface density of the gas at different radii within this cluster using the information of the total mass per pixel found in our Herschel-Planck column density maps.
We compared these measurments with the mass distribution of dense gas obtained from our N$_2$H$^+$ maps (see also Sect.~\ref{DGMF}).
Likewise, we have also estimated the mass surface density of stars from the relative position of the different cluster members in NGC1333, where all these stars are assumed to present a similar mass of 0.5~M$_{\sun}$, as expected for an average star following an standard IMF. 

% Figure 3: surface density
The comparison of the mass surface density from the different components in NGC1333 illustrates, in a quantitative way, some of the properties observed in our maps.
Figure \ref{NGC1333_massdistrib} (upper panel) presents the results obtained for the stellar and gas mass surface density at radii up to 1.3~pc in steps of $\delta$=~0.2~pc from the centre of this cluster. 
From their comparison, it is clear that the distributions of stars and gas are centrally concentrated. However, each of these components present well differentiated central surface densities ($\Sigma_0$) and radial variations. In particular, a steep profile is found describing the distribution of stars with $\Sigma_{stars}\sim$~250 M$_\sun$~pc$^{-2}$ (blue points). Contrary to it, the gas distribution follows a much shallower radial dependency reaching higher central values of $\Sigma_{gas}\sim$~1000 M$_\sun$~pc$^{-2}$ (orange points). 

% Caveats
The radial dependency of the gas and stellar surface densities describes the average properties of the NGC~1333 proto-cluster in a first order approximation. The intrinsic elongated shaped of this cluster, the presence of the previously identified sub-clusters, and its clumpy mass distribution (e.g., see Sect.~\ref{subsec:dense_proto}) hampers the detailed interpreation of the above results. Second-order variations of these absolute values as a function of radius are found depending on the selection of the cluster centre, its eccentricity, and radial extension. Likewise, local variations are expected due to the certainly not homogeneous stellar masses in the case of mass segregation within this cluster. 

Less sensitive than the above absolute measurements, of particular interest is the relative contributions of these gas and stellar components to the total mass of the NGC1333 cluster presented in Fig.~\ref{NGC1333_massdistrib} (mid panel). 
As expected for an embedded proto-cluster during the early stages of evolution \citep{LAD03}, its total mass load is dominated in $\gtrsim$~80\% by its gaseous component at all radii (Fig.~\ref{NGC1333_massdistrib}, upper panel). Compared to it, the relatively smaller mass fraction in stars accounts for $\lesssim$~20\% of the total mass within the same region.

\subsection{Dust thermal structure of the NGC1333 clump}\label{subsec:HP_temp}

% Fig.2, temperature map, Hatchell+
In Fig.~\ref{NGC1333_largescale} (right), we showed the distribution of the dust effective temperatures (T$_{eff}$) within the NGC1333 central clump. The observed values and distribution of T$_{eff}$ are in good agreement with the temperature structure reported by \citet{HAT13} using SCUBA2. Traditionally restricted to the highest column density regions, the larger dynamical range of the new Herschel-Planck maps \citep{ZAR16} provides estimations of the observed T$_{\mathrm{eff}}$ within a range of column densities between 1$^{mag}\lesssim$~A$_V\lesssim~100^{mag}$ with relative errors of $\Delta T_{\mathrm{eff}}/T_{\mathrm{eff}}\lesssim 10\%$. 

% Fig.2, feedback
The observed T$_{eff}$ distribution illustrates the limited impact of the current stellar feedback in the global dust thermal structure of NGC1333.
As deduced from Fig.~\ref{NGC1333_largescale} (right), most of the high column density material of this cloud is found at 
$T_{\mathrm{eff}}\lesssim$~15~K.  Different warm regions are easily recognized for presenting higher temperatures.
The highest temperature measurement in NGC1333 is found at the position of the SVS3 and in the area enclosing BD+30$\degr$549. 
Surrounding them, and with radius of $\sim$200 arcsec, a region of hot dust is observed coincident with the optical reflection nebula produced by these two stars (see left panel). 
Another five hot spots, with sizes of $\sim$50 arcsec, are found at the positions of the different IRAS  sources. These observational signatures demonstrate how some of the most massive objects within this cloud significantly affect the physical conditions of their immediate vicinity. Even in these cases, however, our results suggest that most of the gas in NGC1333 remains insensitive to its emerging stellar population.

 \begin{figure}
   \centering
      \includegraphics[width=\linewidth]{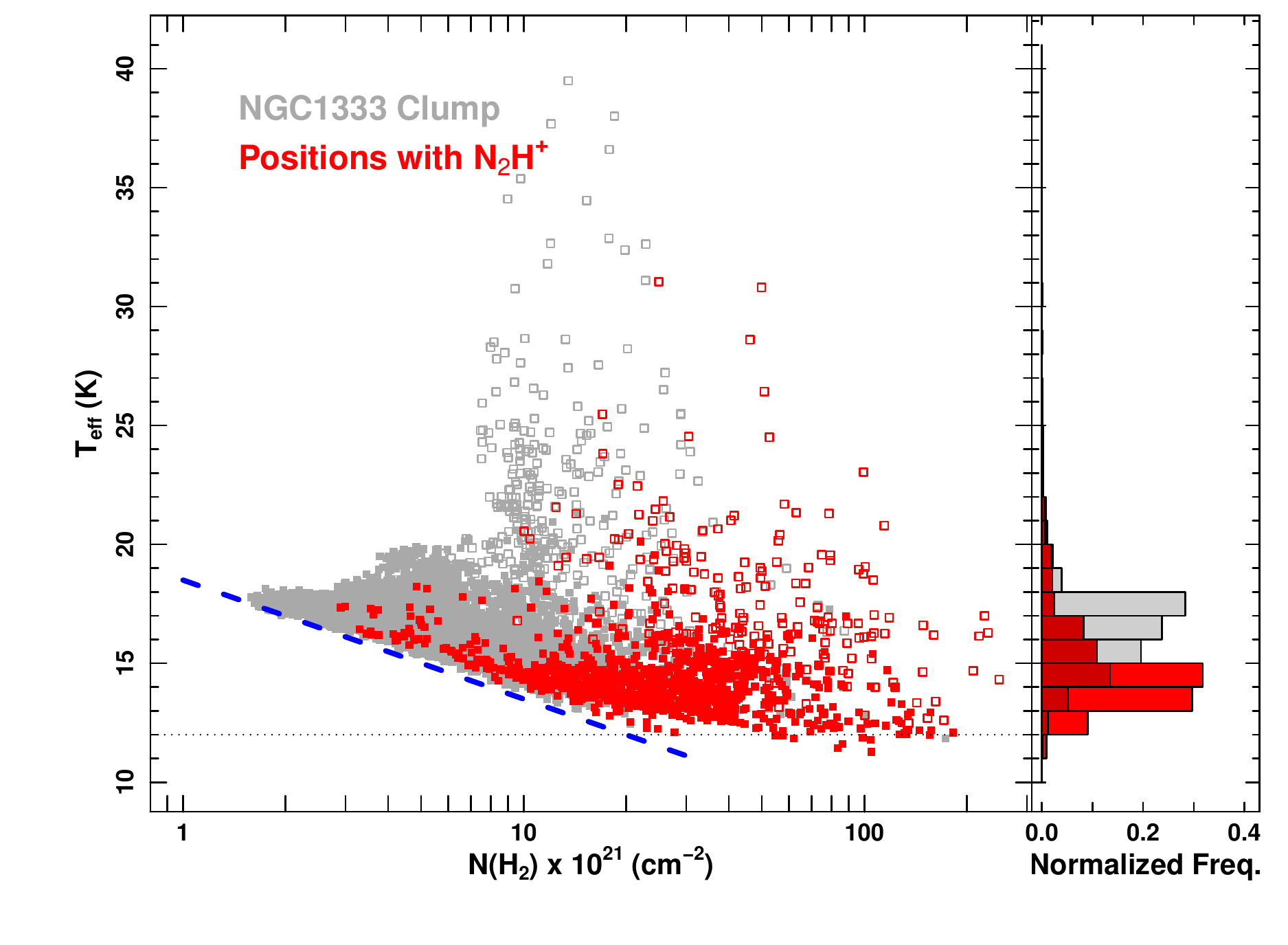}
      \caption{(Left panel) Dust effective temperature (T$_{\mathrm{eff}}$) as a function of the total column density (N(H$_2$); in logarithmic scale) for all the points belonging to the NGC1333 clump presented in Fig.~\ref{NGC1333_largescale} (grey squares). Positions with angular distances $R\le$~200'' from the centre of the optical nebula or $R\le$~50'' from the IRAS 1-8 sources are displayed with open symbols.
      Positions presenting an I(N$_2$H$^+)\ge~1.2 \mathrm{K~km~s}^{-1}$ are highlighted in red. The blue dashed line indicates the eye-fitted lower envelope of all the points in the cloud. The estimated dust effective temperature (T$_{\mathrm{eff}}=$~12~K) floor is indicated by a dotted line. (Right panel) Normalized histograms of the dust effective temperature for the NGC1333 region (grey) compared with only those points having significant N$_2$H$^+$ emission (red).}
             \label{NGC1333_TdustvsNH2}
   \end{figure}

% Fig.3, 
Beyond those regions affected by the internal population of YSOs, we identify an inverse correlation between the dust effective temperature and the total column density N(H$_2$) in the NGC1333 clump. 
The relationship between these two parameters is shown in Fig.~\ref{NGC1333_TdustvsNH2}. 
The highest column density regions within this cloud are always found presenting the lowest dust effective temperatures. Contrary to it, regions at lower column densities exhibit systematically higher temperatures. 
A monotonic decrease of the effective dust temperature is observed at intermediate dust column densities, typically within a range of $\pm$~1.5~K (solid points).

A sharp cut off appears to define a physical limit for the minimum dust effective temperature at a given column density range in Fig.~\ref{NGC1333_TdustvsNH2}.
Empirically, we fit the lower envelope of the observed N(H$_2$)-T$_{\mathrm{eff}}$ values in NGC1333 with a simple linear relationship (blue dashed line in the figure):
\begin{equation}\label{eq:TNfit}
   T_{\mathrm{eff}}[K] = 18.5-5\times log\left(\frac{N(H_2)}{10^{21} [cm^{-2}]}\right).
\end{equation}
The above linear fit describes the global dust properties of this cluster at column densities between 2$^{mag}\lesssim$~A$_V\lesssim~20^{mag}$. 
An inverse correlation between the observed N(H$_2$) and T$_{\mathrm{eff}}$ values is predicted by models of externally heated starless cores \citep[e.g.][]{EVA01} and clouds \citep{BAT15}. In the absence of internal heating sources, this observational correlation reflects the expected outside-in temperature gradient determined by the thermal balance between the external irradiation and the internal self-shielding and dust cooling as a function of the cloud depth \citep[see][for a detailed discussion]{BAT15}.  
As demonstrated in these models, the slope and shape of this correlation depends on the detailed physical structure of the cloud and the intensity of the heating source, typically assumed as the Interstellar Radiation Field (ISRF). 
Based on a qualitative comparison with our observational results, we conclude that the internal thermal structure of the NGC1333 proto-cluster is primarily determined by the external irradiation of the cloud with a minor contribution of its embedded stellar population at large scales.

%%%%%%%%%%%%%%%%%%%%%%%%%%%%
%% Dense tracers
%%%%%%%%%%%%%%%%%%%%%%%%%%%%
\section{Molecular tracers: dense gas properties}\label{sec:densegas}

% Molecular emission and dense tracers
Unlike the dust continuum emission, the molecular emission of the gas is strongly affected by opacity, excitation, and chemical effects.
Although intrinsically more difficult to interpret, the emission of different key tracers can be used to selectively study different gas properties in molecular clouds.
That is the case of N-bearing molecules like N$_2$H$^+$ and NH$_3$.
While their formation is inhibited in the diffuse gas, both N$_2$H$^+$ and NH$_3$ molecules rapidly increase their abundances after the CO 
	freeze-out onto the dust grains at temperatures $<$~20~K above
and at
densities n(H$_2$)~$\gtrsim$~10$^{4}$ cm$^{-3}$ \citep[see ][for a review]{BER07}. With favourable excitation conditions at these densities, some of the ground level transitions of these molecules like the N$_2$H$^+$ (1--0) and NH$_3$ (1,1) and (2,2) lines are regularly employed on the study of dense cores \citep[e.g.][]{MYE83,CAS02}. In this work, we have used large scale emission maps of these two molecules to investigate the properties of the dense gas along the entire NGC1333 region.
 
\subsection{N$_2$H$^{+}$ vs. NH$_3$ line emission: twin dense tracers}\label{subsec:Nbearing}

   \begin{figure*}
   \centering
   \includegraphics[width=\textwidth]{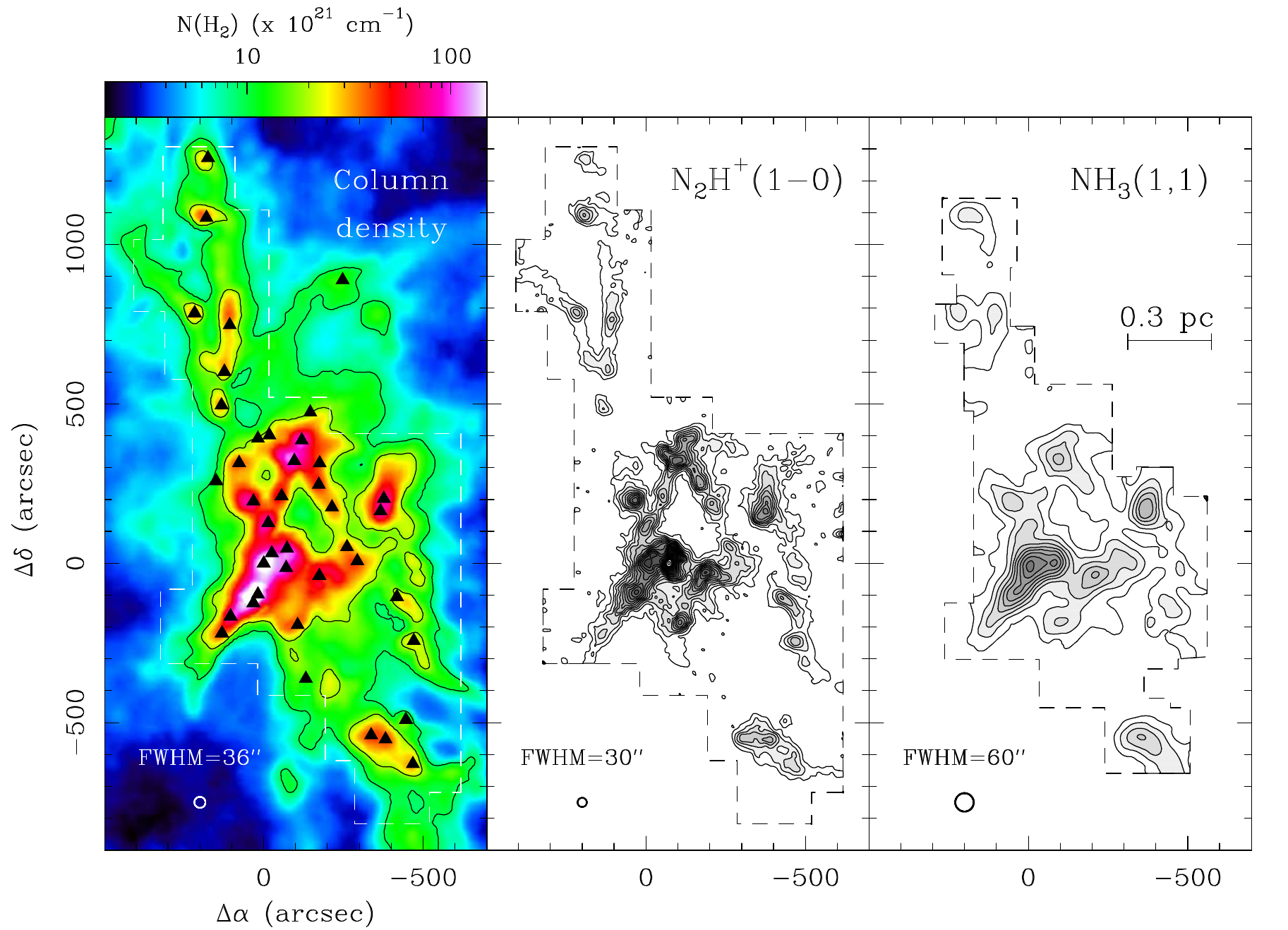}
      \caption{(Left) Total gas column density towards the NGC1333 central clump. Contours similar to Fig.~\ref{NGC1333_largescale}. The solid triangles indicate the position of the different dense cores identified by \citet{ROS08}. (Centre) N$_2$H$^{+}$ (1--0) integrated emission map obtained with the IRAM30m telescope. (Right) NH$_3$ (1,1) integrated emission map obtained with the Effelsberg 100m telescope. 
		Offsets are referred to the map centre with coordinates $(\alpha,\delta)_{J2000} = (03^h29^m08^s.9,+31\degr15'12'')$ in radio projection.
      Contours are equally spaced every 1.2 K~km~s$^{-1}$ and 1.0 K~km~s$^{-1}$ in the N$_2$H$^{+}$ and NH$_3$ maps, respectively. In each case, the area surveyed by our molecular observations is enclosed by a dashed line. The area mapped in N$_2$H$^{+}$  is also delineated by a white dashed line on the left panel.
      The different FWHM are indicated at the lower left corner in the corresponding plots.
      }
             \label{NGC1333_densetracers}
   \end{figure*}

Figure~\ref{NGC1333_densetracers} shows the total integrated intensity maps of both N$_2$H$^+$ (1--0) (central panel) and NH$_3$ (1,1) lines (right panel) in comparison to the total column density of gas (left panel) along the NGC1333 clump. 
\citet{ROS08} compiled the most recent census of dense cores in Perseus, including NGC1333, combining different surveys in both lines \citep{JIJ99} and continuum \citep{KIR06,ENO06}. 
Using GBT observations with a resolution of 31~arcsec, these authors surveyed  a total of 42 dense cores in ammonia within the area mapped by our IRAM30m observations. Among them, 37 ($\sim$~88\%) exhibit emission counterparts in our N$_2$H$^{+}$ (1--0) and NH$_3$ (1,1) maps, typically coincident with the positions of local maxima in total gas column density (solid black triangles). Corresponding to marginally detections in the Rosolowsky et al sample, the other 5 positions (Cores 54, 61, 71, 76, and 85) appear at column densities of A$_V<$~15$^{mag}$ with no clear compact detection in the molecular emission nor at the continuum. 
Similarly, prominent detections are found at the position of all the N$_2$H$^{+}$ clumps identified by \citet{WAL07} using interferometric BIMA+FCRAO observations at a resolution of 10~arcsec. Dilution and blending effects in our single-dish beam, as well as the combination of extended emission and spatial filtering effects, complicate a direct comparison with this last sample. 

Beyond these dense cores, our molecular maps also reveal large-scale emission in both N$_2$H$^{+}$ and NH$_3$ tracers not detected in previous surveys in NGC1333. 
Most of the peaks detected in these molecular maps are connected by a prominent and extended emission clearly detected down to A$_V\sim$~8$^{mag}$. The widespread emission of these N-bearing molecules indicate the presence of extended depletion effects at cloud scales.  This global chemical evolution suggests the presence of large amounts of dense gas within this region. 

   \begin{figure}
   \centering
   \includegraphics[width=\hsize]{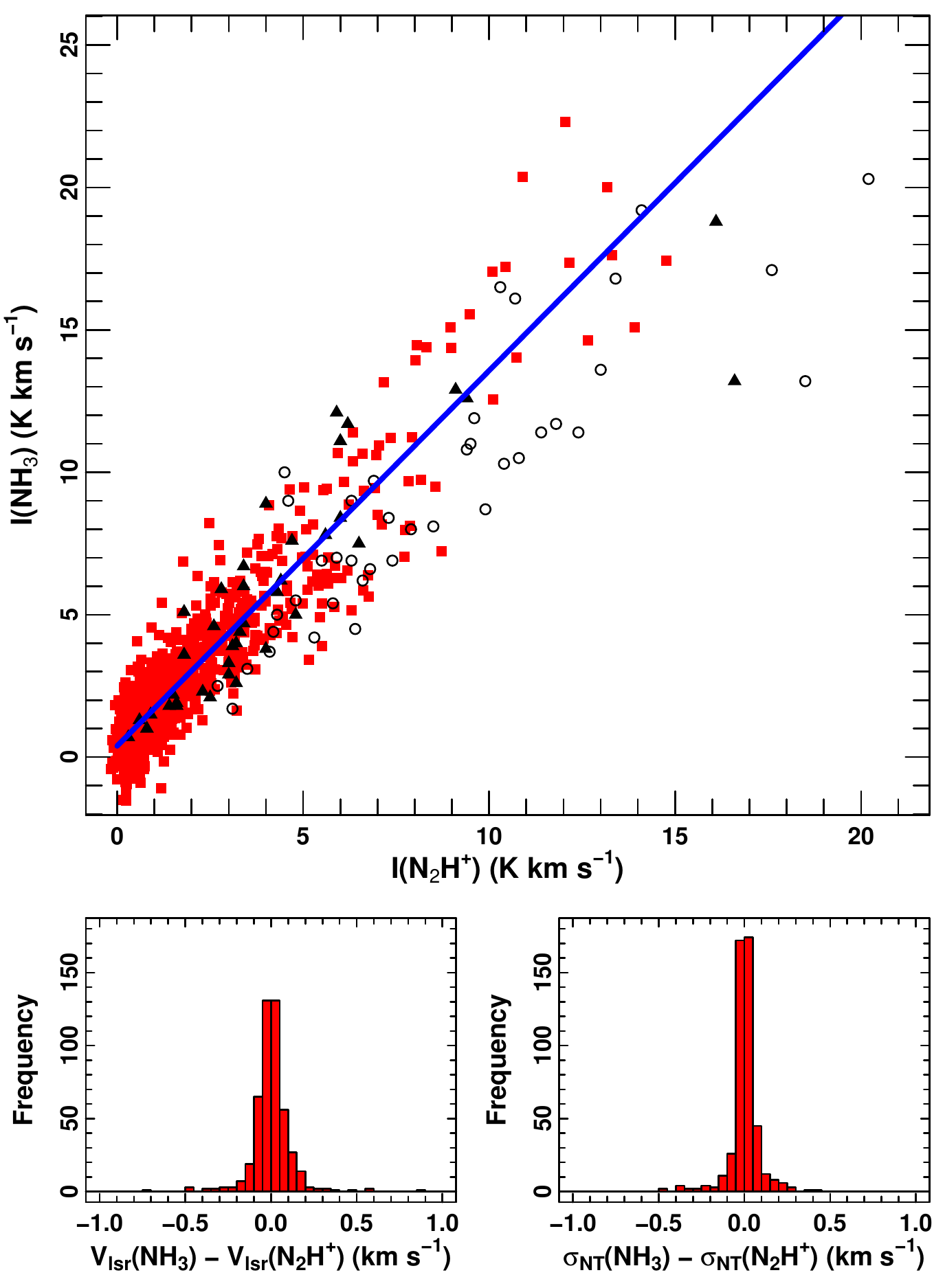}
      \caption{ (Upper panel) Pixel-to-pixel comparison between the N$_2$H$^{+}$ (1--0) and NH$_3$ (1,1) integrated emission within NGC1333, both convolved to 60 arcsec. The blue line denotes the linear fit described by Eq.~\ref{eq:N2Hp_NH3}. The results obtained in Perseus by \citet{JOH10} for both prestellar (solid triangles) and protostellar cores (open circles) are superposed in black. (Lower subpanels) Histograms of the differences in velocity (left) and non-thermal velocity dispersion (right) between N$_2$H$^{+}$ and NH$_3$ for all the positions fitted with a S/N~$\ge$~3. For positions with multiple lines, only the comparison to the nearest component in velocity is displayed (see Sect.~\ref{sec:kinematics} for a discussion about the gas kinematics).
      }
             \label{NGC1333_N2H+_vs_NH3}
   \end{figure}

From the comparison of the emission properties of both N$_2$H$^{+}$ (1--0) and NH$_3$ (1,1) lines in a sample of 71 pre- and protostellar dense cores in Perseus, \citet{JOH10} concluded that the formation and destruction of these two N-bearing molecules are strongly coupled at high density regimes. This parallel evolution is already evident in the total gas content traced by these molecules in NGC1333. 
In Figure~\ref{NGC1333_N2H+_vs_NH3} (upper panel), we compare the line emission properties of our N$_2$H$^{+}$ (1--0) and NH$_3$ (1,1) maps, both convolved to a common resolution of 60~arcsec (red squares). 
In close agreement with the results of \citet{JOH10} (black symbols), the integrated intensities of these two molecules present a tight and linear correlation along the entire cloud (blue line) :
\begin{equation}
   I~[NH_3 (1,1)] = 1.32\cdot I~[N_2H^{+} (1-0)] + 0.39 \label{eq:N2Hp_NH3}
\end{equation}
Their similarities can also be extended to the kinematic properties of these two lines (lower subpanel). 
For the gas traced by these two species, we find average differences within less than 1/5 of our spectral resolution comparing both their central velocities (lower left) and non-thermal velocity dispersions (lower right).
Similarly to the results obtained in dense cores \citep[e.g.,][]{PAG09}, the strong correlation between their observational properties demonstrates that these two molecular species act as twin tracers of the dense gas content in NGC1333,  at least at the scales resolved by our single-dish observations.

The detection of emission of both N$_2$H$^{+}$ and NH$_3$ tracers is typically interpreted as function of their distinct critical densities. In this framework, the NH$_3$ (1,1) emission is assumed to be sensitive to the gas densities above $\sim$~10$^3$~cm$^{-3}$, while the N$_2$H$^{+}$ (1--0) line is meant to be excited only at densities $\gtrsim$~10$^5$~cm$^{-3}$. Contraty to it, our observations show identical emission properties for these two tracers, in agreement with previous results \citep{JOH10}.
These similarities indicate that the detection of these two molecules is controlled by their formation mechanism and chemically triggered by the CO freeze-out ocurring in the gas reaching densities $\gtrsim  5\times 10^4$~cm$^{-3}$ (see also Sects.~\ref{subsec:gastemp} and \ref{subsec:densegas}).

Our findings allow us to directly combine the information independently provided by each of these tracers for the study of NGC1333. In particular, we have used different measurements of the NH$_3$ lines to estimate the gas kinetic temperature of the dense gas component of this cluster (Sect.~\ref{subsec:gastemp}). On the other hand, the higher resolution and sensitivity of our N$_2$H$^{+}$ spectra made them the best choice for the study of both the mass distribution (Sect.~\ref{subsec:densegas}) and gas kinematics (Sect.~\ref{sec:kinematics}) within this region.

\subsection{NH$_3$ vs. Continuum: thermal gas-to-dust coupling}\label{subsec:gastemp}

   \begin{figure}
   \centering
   \includegraphics[width=\hsize]{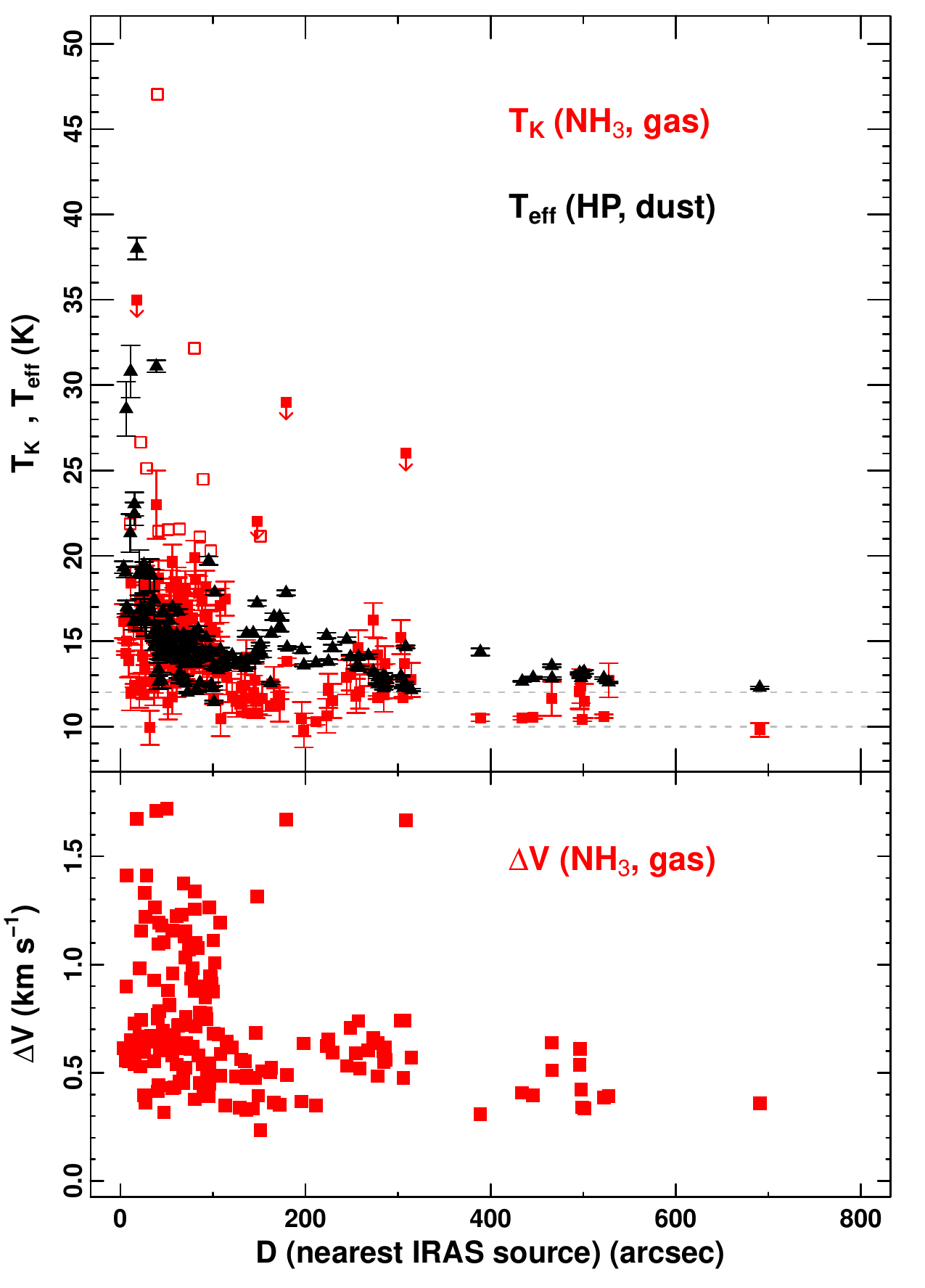}
      \caption{(Upper panel) Gas kinetic (T$_K$; red squares) and dust effective (T$_{eff}$; black triangles) temperatures as a function of the projected radius to the nearest IRAS source within NGC1333. The plot combines all positions with derived T$_K$ values in this work together with those surveyed by \citet{ROS08}. The bars indicate the 1$\sigma$ errors of both gas and dust temperatures, respectively. The red open squares indicate positions with T$_K>$~20~K values (see text). The arrows indicate positions with T$_K$ values listed as upper-limits by \citet{ROS08}. (Lower panel) Measured ammonia linewidths ($\Delta V$) within the same positions.}
             \label{NGC1333_Tk}
   \end{figure}
   
   Taking advantage of their favourable observational and physical properties, the analysis of the two fundamental ammonia (1,1) and (2,2) inversion transitions is routinely employed as gas thermometer \citep[see][for a review]{HO83}. In addition to the previous point-like surveys in NGC1333, we have investigated the thermal properties of the dense gas content detected within this proto-cluster. We have combined the results of all the NH$_3$ (1,1) and (2,2) components fitted with S/N$\ge$~3 (see Sect.~\ref{sec:kinematics}) to, first, derive the ammonia rotational temperature and, then, the gas kinetic temperature (T$_K$) using standard techniques \citep{BAC87,TAF04}. 
   
Values of the gas kinetic temperature T$_K$ were obtained in 121 positions along the NGC~1333 clump. Typically limited by the detection of the NH$_3$ (2,2) lines, most of these measurements are concentrated towards the high column density regions that surround the IRAS 2, 3, 4, and 6 sources. 
Within one beam, 19 of these detections coincide with the dense cores surveyed in ammonia by \citet{ROS08} using dedicated GBT-observations. Despite the different sensitivities and methods used in this last study, our T$_K$ estimates agree within $\sim$~1~K with those obtained by Rosolowsky et al for T$_K<$~20~K  
and  are assumed as the typical errors in our estimates.
Those T$_K>$~20~K derived from our data are only considered as indicative of warmer temperatures. Additional ammonia transitions would be necessary to constrain these higher temperatures \citep[see][for a discussion]{TAF04}.

Similar to the effective dust temperatures (Sect.~\ref{subsec:HP_temp}), 
we observe a monotonic increase of the ammonia derived gas kinetic temperature T$_K$ in the proximity of the most prominent IRAS objects. The influence of these strong heating sources is illustrated by the dependency of T$_K$ as a function of the distance D to the nearest IRAS source in Figure~\ref{NGC1333_Tk} (upper panel).  
We measured a temperature floor of $\sim$~10-12 K for those dense regions far from the active sites of star formation (D$\gtrsim$~150 arcsec). In contrast, and at D~$\lesssim$~50-100 arcsec, the strong radiation emerging from these embedded sources efficiently heats up the gas above T$_K>$~15~K. These heating effects are translated into an increasing line thermal broadening, clearly observed in our ammonia linewidth measurements (lower panel). 
Second-order effects are explained by the proximity to the warm edges of the optical reflection nebula.

Figure~\ref{NGC1333_Tk} (upper panel) also includes the effective dust temperature T$_{eff}$ measurements at the same positions of our ammonia observations (black triangles). At large distances from the IRAS heating sources, T$_{eff}$ exhibits a roughly constant temperature of $\sim$~12-13~K, typically $\sim$~2~K higher than their corresponding gas kinetic temperatures.
Similar temperature differences have been previously reported by \citet{FOR15} comparing these two observables in a series of starless cores in the Pipe Nebula.
As discussed by these authors, the observed correlation between these T$_{eff}$ and T$_{K}$ values are likely explained by the expected dust-gas thermal coupling at the densities traced by ammonia \citep[i.e., $>10^4$~cm$^{-3}$][]{GOL01}.
In NGC1333, this parallel evolution is altered at distances D~$<$~100 arcsec form the most prominent heating sources within this region and when both gas and dust temperatures rise above $\gtrsim$~15~K.
This apparent thermal decoupling could be produced by the independent heating and cooling mechanism of these gas and dust components and the increase of the UV irradiation in proximity of the newborn YSO as observed in more massive regions \citep[e.g.,][]{BAT14,KOU15}. Investigating the nature of these mechanisms remains, however, unclear from our data. 
Despite these local effects, the derived Herschel-Planck dust effective temperatures appear as a reliable proxy of the gas kinetic temperatures in the densest regions of the NGC1333 proto-cluster. 

\subsection{N$_2$H$^{+}$ vs. Continuum: cold and dense gas}\label{subsec:densegas}

   \begin{figure}
   \centering
   \includegraphics[width=\hsize]{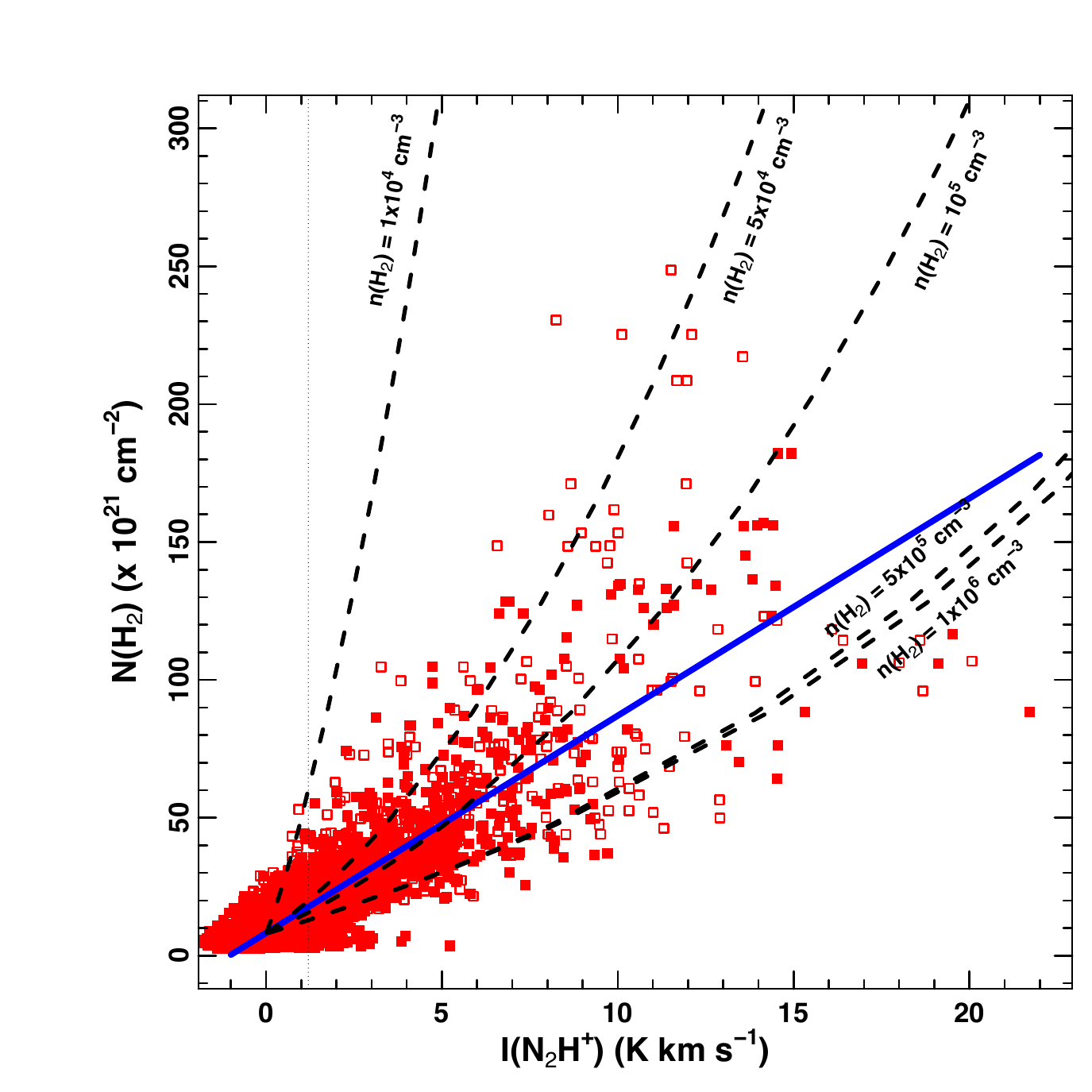}
      \caption{Comparison between the N$_2$H$^+$ (1--0) integrated intensity (i.e., I(N$_2$H$^+$)) and H$_2$ column density (i.e., N(H$_2$)) in NGC1333. The vertical dotted line indicates the intensity threshold defining the first contour of the emission map in Fig.~\ref{NGC1333_densetracers}. The linear fit describing Eq.~\ref{eq:N2Hp_H2} is displayed by a blue solid line. Symbols are similar to those in Fig.~\ref{NGC1333_TdustvsNH2}. RADEX radiative transfer calculations for N$_2$H$^+$ (1--0) emission at densities n(H$_2$)~$=[0.1, ~0.5, ~1.0, ~5, ~10]\times10^5$~cm$^{-3}$ are overploted using black dashed lines (see text).  }
             \label{NGC1333_coldensity}
   \end{figure}

A careful inspection of Fig.~\ref{NGC1333_densetracers} shows that the two N$_2$H$^{+}$ (1--0) and NH$_3$ (1,1) emission maps do not only present a parallel evolution but also  strong similarities with the continuum emission. 
Indeed, the variations and maxima on the N$_2$H$^{+}$ integrated emission typically mimic the distribution of the total column density map at A$_V>$~10$^{mag}$.
This correlation can be quantified in a pixel-by-pixel comparison like the one shown in Fig.~\ref{NGC1333_coldensity}. 
To produce this plot, we have compared the integrated N$_2$H$^{+}$ emission in our molecular spectra with the nearest position surveyed by our Herschel-Planck continuum maps.
The N$_2$H$^{+}$ integrated emission linearly increases with the total column density of gas traced in the  continuum. From a linear fit of all the points in this plot, we obtain an empirical relationship between these two observables following:
\begin{equation}
   \centering
   N~(H_2) [cm^{-2}] = \left(7.9\cdot I~[N_2H^{+}]~[K~km~s^{-1}]+8.2\right)\times 10^{21}
     \label{eq:N2Hp_H2}
\end{equation}
Deviations from this linear relation are found in the vicinity of deeply embedded sources and the optical nebula (open squares) likely due to the strong temperature gradients generated within these regions (see Sect.\ref{subsec:gastemp}). With perhaps the exception of these positions, Eq.~\ref{eq:N2Hp_H2} satisfactory reproduces the observed correlation between the N$_2$H$^{+}$ emission and the total column density up to values of N(H$_2$)~$\sim 150\times10^{21}$~cm$^{-2}$ with variations of less than a factor of 2.

A linear and tight correlation is expected if the N$_2$H$^{+}$ emission remains optically thin and its abundance plus excitation conditions do not change significantly within the range of column densities sampled by our single-dish observations \citep[e.g.,][]{JOH10}. 
\citet{TAF15} demonstrated that, at the densities where the N$_2$H$^{+}$ (1--0) line is effectively produced (i.e., n(H$_2$)~$\gtrsim 10^4$~cm$^{-3}$), the N$_2$H$^{+}$ conversion factor presents a roughly constant value of N(N$_2$H$^{+}$)$\sim 2 \times 10^{12}$~I[N$_2$H$^{+}$]. Combined with Eq.~\ref{eq:N2Hp_H2}, this last relationship leads to a typical abundance of X$_0$=X(N$_2$H$^{+}$/$H_2$)~$\sim 2.5 \times 10^{-10}$, in good agreement with previous estimates in dense cores \citep{CAS02,TAF02}. We have tested these predictions using the radiative transfer code RADEX \citep{VDT07} assuming a constant kinetic temperature of T$_K=$~10~K (Sect.~\ref{subsec:gastemp}), homogeneous X$_0$ abundance (see above), and a column density threshold of $8.2\times 10^{21}$~cm$^{-2}$ similar to Eq.~\ref{eq:N2Hp_H2}. According to these models, the observed dependency between the N$_2$H$^{+}$ (1--0) line intensities with the total gas column density in NGC1333 corresponds to the expected variations for a gas at densities between $> 5\times10^4$ and  $\sim10^6$~cm$^{-3}$ (black dashed lines).  

The empirical fit defined by Eq.~\ref{eq:N2Hp_H2} suggests a minimum column density value required for the formation of dense gas.
This threshold is estimated from the value of the abscissa obtained in this linear correlation with a value
 of $\sim 8.2\times10^{21}$~cm$^{-2}$ (or A$_V\sim 9^{mag}$ of visual extinction). As highlighted in Fig.~\ref{NGC1333_TdustvsNH2} (red squares), only 4\% of the positions with detected N$_2$H$^{+}$ emission are found at lower column densities than this value.   
Above this threshold, and excluding those positions likely affected by the stellar feedback (open symbols), 80\% of those dense regions with N$_2$H$^{+}$ detections are found presenting dust effective temperatures of T$_{eff}<$16~K, assumed as an upper limit of the true gas kinetic temperature (see Sect.~\ref{subsec:gastemp}).
According to these findings, the dense gas traced by our N$_2$H$^{+}$ and NH$_3$ observations is not only 
found at the highest extinctions in the NGC1333 proto-cluster,
but also originates from its densest (i.e., n(H$_2$)~$>5\times10^4$ cm$^{-3}$) and coldest (T$_K\lesssim$~15~K) gas component within this cloud.

\subsection{Total mass and dense gas mass fraction}\label{DGMF}

% Mass
We have derived the total amount of dense gas within the NGC1333 clump 
using (only) the slope of Eq.~\ref{eq:N2Hp_H2} defining the
 conversion factor between the observed N$_2$H$^{+}$ intensities and the equivalent N(H$_2$) column densities. 
Adding the contribution of all of the positions within the first contour of the maps presented in Fig.~\ref{NGC1333_densetracers}, we obtained a total of $\sim$~250~M$_\sun$ of dense gas within the area mapped by our IRAM~30m observations. 

% Completness
Due to the limited coverage of our N$_2$H$^{+}$ maps compared to the total extension of the NGC1333 ridge, it is important to quantify the completeness of this last measurement. 
As illustrated in Fig.~\ref{NGC1333_TdustvsNH2}, the detection of dense gas is favoured towards high column densities and low dust effective temperatures. 
Within the area covered by our IRAM30m observations, 80\% of the points with T$_{eff}\le$~16~K and A$_V\ge 20^{mag}$ are found with I(N$_2$H$^{+}$)~$\ge$~1.2~K~kms$^{-1}$. The mass contained within our molecular map boundaries represents $\sim$~95\% of the total mass at similar dust effective temperatures and column densities along this entire ridge (Sect.\ref{sec:NGC1333}). From these estimations, we expect our molecular observations to recover most of dense gas content of the entire NGC1333 region.

% Radial distribution
Similar to the total gas and stellar distributions (Sect.\ref{subsec:HP_mass}), the dense gas in NGC1333 proto-cluster is highly concentrated towards its centre.
This behaviour is evidenced in the mass surface density of dense gas presented Fig.\ref{NGC1333_massdistrib} (upper panel; red dashed line), directly calculated from our N$_2$H$^{+}$ integrated intensity maps.
In relative terms, the dense gas mass fraction also varies as a function of the cluster radius. As shown in Fig.~\ref{NGC1333_massdistrib} (mid panel; red dashed line), the dense gas accounts for $\gtrsim$~40\% of the total mass at R~$\lesssim$~0.3~pc compared to the total mass load (Stars+Gas; blue line) within the same region. This fraction slowly decreases below $<$~20\% at R~>~0.7~pc due to the lack of a significant pockets of dense gas at larger radii. Although representing a relative low fraction of the total mass load at large scales, these numbers demonstrate the dominant role of this dense gas component within the central clump of NGC1333.

\subsection{Direct correlation between protostars and dense gas: evolutionary timescales and efficiencies}\label{subsec:dense_proto}

   \begin{figure*}
   \centering
   \includegraphics[width=\hsize]{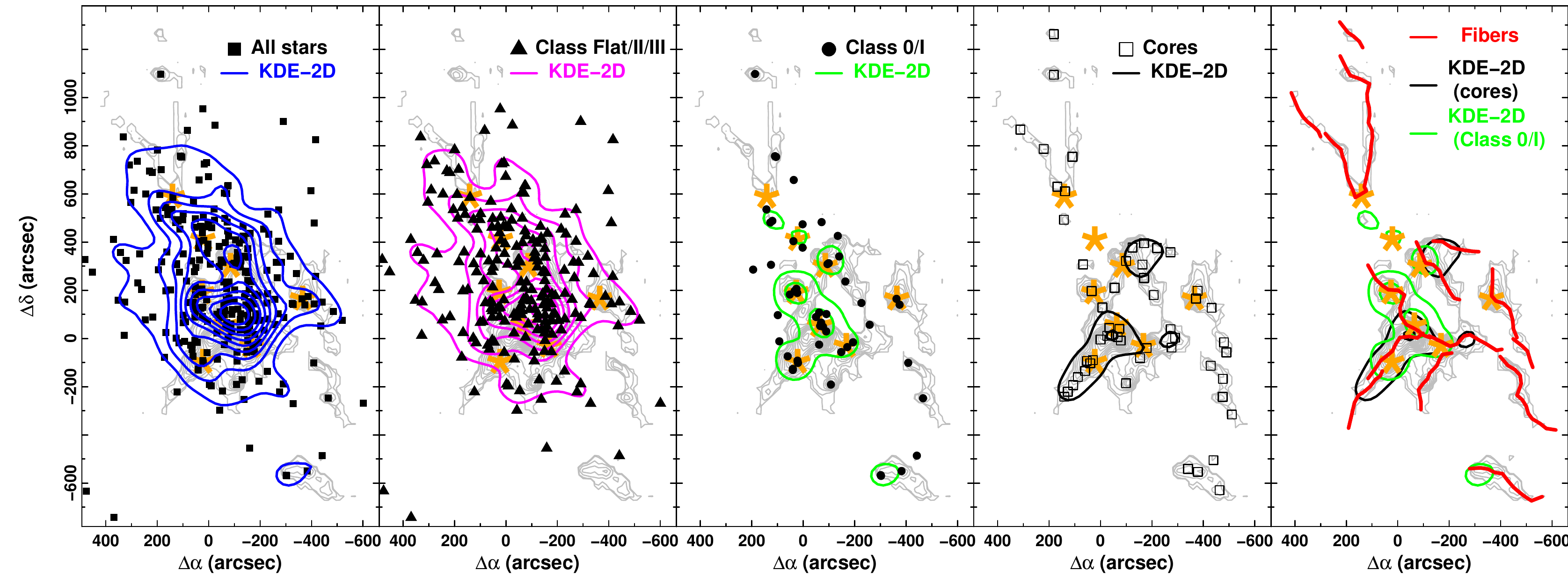}
      \caption{Spatial distribution and surface density of stars and gas within NGC1333 obtained using 2D KDEs with a FWHM~=~220~arcsec. From left to right: (1) All stars (solid squares); (2) Classes Flat, II, and III (solid triangles); (3) Classes 0/I (solid circles); (4) Dense cores (open squares); (5) Fiber axis (red lines). Contour levels are equally spaced every 100 objects pc$^{-2}$ in all submaps. For reference, the N$_2$H$^{+}$ emission (grey contours) as well as the positions of the different IRAS sources  (orange stars) are indicated in all subpanels.}
             \label{fig:NGC1333_KDE2D}
   \end{figure*}

% Previous studies
Several studies have attempted to connect different indicators of star formation and the gas properties in molecular clouds.
In the case of Perseus, \citet{HAT05} demonstrated that the probability of finding a dense core rapidly increases with column density. Independent results have pointed out that both protostellar cores \citep{KIR06,ENO06} and embedded protostars \citep{JOR07} are typically found in regions with column densities A$_V>$~5$^{mag}$.
More recently, \citet{SAD14} associated the location of Class 0 objects to the highest column densities in these cloud.
Our new observations allow us to go one step further studying not only the correlation between these young objects and total gas column density but also their connection to the position and distribution of dense gas in NGC1333.

% KDE 
The position and location of the two North and South subclusters in NGC1333 (see also Sect.~\ref{NGC1333_massdistrib}) are easily identified from the surface density of objects presented in Figure~\ref{fig:NGC1333_KDE2D} (first panel). These plots are obtained using a 2D gaussian kernel density estimate (KDE), with a kernel full with half maximum FWHM=220~arcsec, for all the clusters members identified by \citet{REB15}.
From a study of the spatial distribution of the different stellar populations in NGC1333, \citet{GUT08} concluded that this substructure is primarily determined by the more evolved Class II/III objects within this region. On the contrary, the younger protostars appear to follow the filamentary structure traced in previous continuum observations. 
These results are confirmed by different surface density estimates presented in Figure~\ref{fig:NGC1333_KDE2D} (second and third panels) in comparison with the dense gas emission traced in N$_2$H$^{+}$ (grey contours in these plots). 
As discussed in Sect.~\ref{subsec:kinematics:fibers}, this primordial stellar structure is likely imprinted by the initial distribution of the dense gas arranged forming a complex fiber network (last panel).

\begin{figure}
   \centering
   \includegraphics[width=\hsize]{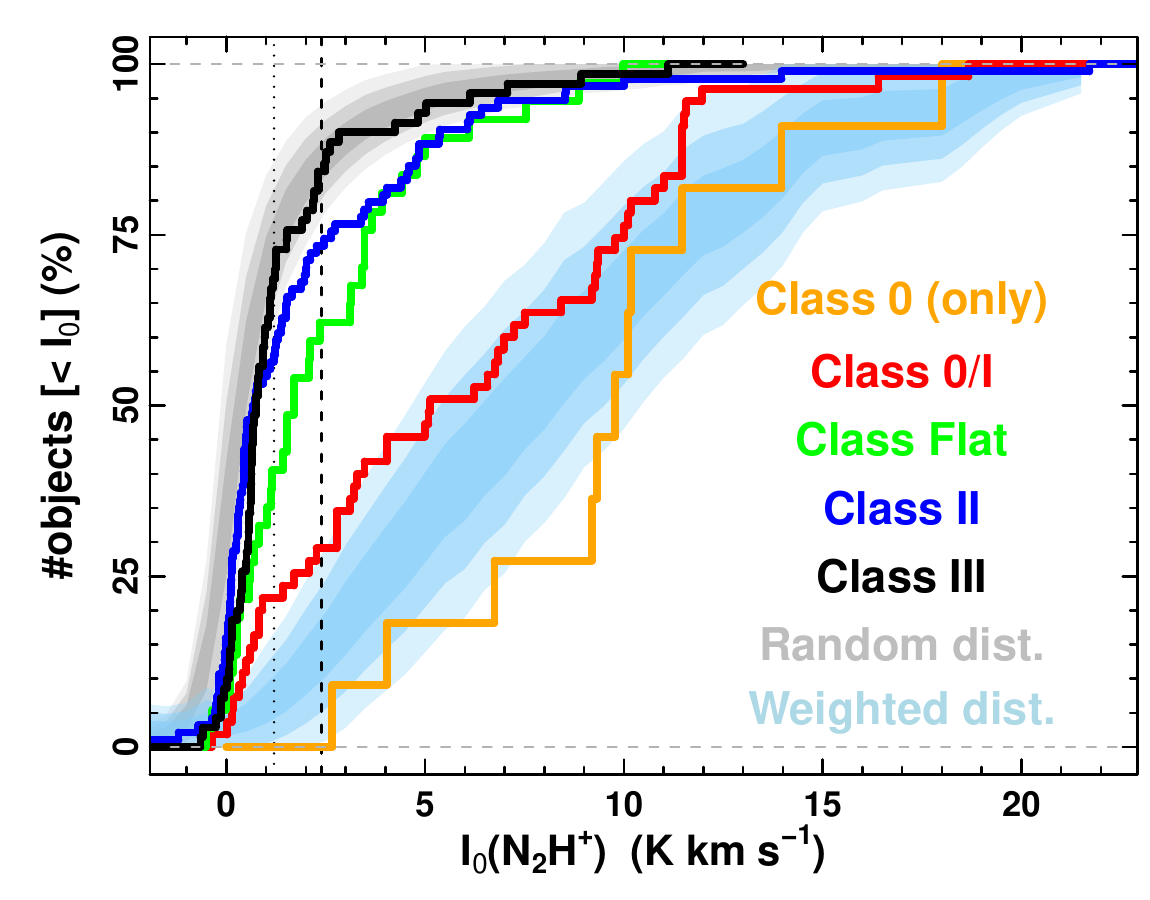}
      \caption{Cumulative fraction of stars as a function of the N$_2$H$^{+}$ intensity (I$_0$). The distribution of Class 0 (orange), Class 0/I (i.e., all protostars; red), and Classes Flat (green), II (blue), and III (black) are independently displayed. The two shaded areas (with 1$\sigma$, 2$\sigma$, and 3$\sigma$ contours) correspond to the cumulative fraction of objects resulting from random distribution of objects (grey) and a distribution where the probability of finding an object is proportional to the second power of the N$_2$H$^{+}$ intensity (light blue) (see text). The two vertical lines define the N$_2$H$^{+}$ intensity of the first (dotted line) and second (dashed line) contours 
      	 in Fig.\ref{NGC1333_densetracers}, respectively.}
             \label{YSOs_N2Hp}
   \end{figure}

% Cumulative plot: presentation
To quantify the correlation between dense gas and stars in NGC1333, in Fig.~\ref{YSOs_N2Hp} we present the cumulative distribution of objects as a function of the N$_2$H$^{+}$ intensity. To construct this plot, we considered only those objects inside the area surveyed by our IRAM~30m observations assigning to each star the N$_2$H$^{+}$ intensity value of the nearest pixel in our molecular maps. A visual inspection of this figure reveals strong differences comparing the distinct stellar classes inside this cloud. All Class 0 objects (orange line) are found at positions containing high N$_2$H$^{+}$ intensities, with a minimum value corresponding approximately to the second intensity contour in Fig.~\ref{NGC1333_densetracers} (dashed line). This behaviour is also followed by all the protostars (red), including both Class 0/I objects, with more than 70\% of these objects located within the same intensity levels. This trend systematically varies as a function of the stellar classes, indicative of their ages. Only 40\% in the case of Flat sources and $\sim$~25\% of the Class II and III objects are found within these high intensity contours. These differences are highlighted by comparing the distribution of stars outside the first emission contour of our N$_2$H$^{+}$ maps (vertical dotted line). While only $\sim$~15\% of the Class 0/I objects are located in areas devoid of N$_2$H$^{+}$ emission, this number rises up to $\sim$~60\% in the case of Class II/III stars. 

% Comparison to models: Class II/III
The results obtained in Fig.~\ref{fig:NGC1333_KDE2D} suggest not only that the different population of stars are not equally distributed but also that they present distinct degrees of correlation with respect to the dense gas in NGC1333.
We have compared the observed cumulative fraction of Class II/III objects with the statistical results obtained after 1000 simulations where the same number of objects are distributed inside our IRAM30m N$_2$H$^{+}$ maps following a random distribution in position with uniform probability.
As shown in Fig.~\ref{YSOs_N2Hp} (grey contours), these random models reproduce the distribution of Class II/III objects in NGC1333. 
These similarities indicate that most (if not all) of the Class II/III YSOs found inside this region are not related to the dense gas currently observed in this cloud. Instead, they suggest that the correlation between some of these objects and regions with high N$_2$H$^{+}$ emission is likely caused by a chance superposition along the line-of-sight.

% Comparison to models: Class 0/I
Contrary to these more evolved YSOs, the Class 0/I protostars do exhibit a strong correlation with the dense gas structure within this cloud. Using the same type of analysis, we have run a new series of 1000 models for the distribution of a similar number of Class 0/I objects where, in this last case, we set the probability of finding one of these objects to be proportional to the second power of the N$_2$H$^{+}$ emission in our maps. As denoted by their comparison in Fig.~\ref{YSOs_N2Hp}, the distribution of protostars closely follow the typical values found for these weighted models (light blue contours). From both their spatial distribution and their point-to-point correlations, we conclude that only the youngest Class 0/I protostars are directly related to the dense gas detected in N$_2$H$^{+}$.

% Timescales
The above results impose an upper limit for the evolutionary timescales of dense gas in the NGC1333 proto-cluster.  Particularly, the direct match found between the position of the Class 0/I protostars and actual distribution of dense gas detected in N$_2$H$^{+}$, in addition to the lack of a similar correspondence with the more evolved Class Flat/II/III objects, indicates that this gas component should evolve within a characteristic timescale comparable to the typical formation timescale of a Class I object, that is, $\sim$~0.5~Myr \citep{EVA09}. With a free-fall time of  $\tau_{ff}\sim$~0.11~Myr for n(H$_2$)~$\sim 10^{5}$ cm$^{-3}$ (Sect.~\ref{subsec:densegas}), we then estimate that a characteristic time of $\tau_{dense}\gtrsim 5 \times \tau_{ff}$ for the dense gas within the NGC1333 proto-cluster.
Although relatively large compared to its free-fall time, the lifetime derived for the dense gas evolution is shorter than the total lifetime of the NGC1333 cluster and, in particular, the first generation of stars formed inside this region, nowadays observed as evolved Class II/III objects ($\sim$~2~Myr or $\sim 20\tau_{ff}$). 
These results suggest a dynamical scenario where similar dense gas structures have been continuously generated and dispersed along the evolution of this cluster.

% SFE and SFR
A fundamental parameter describing the evolution of young stellar clusters is the fraction of gas mass converted into stars, namely, the Star Formation Efficiency (SFE) \citep{LAD03}. We have defined the current SFE in dense gas as SFE(dense)~$=\frac{M_{proto}}{M_{proto}+M_{dense}}$, 
where M$_{proto}$ and M$_{dense}$ correspond to the mass in protostars (i.e., Class 0/I) and the total mass detected in N$_2$H$^{+}$, respectively. For a typical stellar mass of 0.5 M$_{\odot}$, so M$_{proto}=N_{proto} \times 0.5$~M$_{\odot}$, we estimate current values of SFE~$\sim$~10\% 
for the dense gas within NGC1333 at densities of $\sim 10^{5}$ cm$^{-3}$. This SFE value remains constant and independent of the position within the cluster (within the estimated errors assuming Poisson statistics for the stellar counts), as indicated by the calculations of this parameter at the different radial bins presented in Fig.~\ref{NGC1333_massdistrib} (lower panel).
This result is in apparent disagreement with the increasing number of young protostars found towards the cluster centre suggesting a higher SFE (green line; upper panel). On the contrary, our observations demonstrate that this effect is directly explained by a similar increasing amount of dense material within the same regions (red line; upper and mid panels) collapsing and forming stars with constant efficiency.

\section{Dense gas kinematics: fibers in NGC1333}\label{sec:kinematics}

As demonstrated in previous sections, the origin of the cores and protostars of NGC1333 is directly connected to the distribution and properties of its dense gas content. Investigating the dynamical state of this gaseous component is of paramount importance to the star formation mechanism within this proto-cluster. 
The dense gas kinematics within the central R$\lesssim$~0.3~pc of the NGC1333 cluster has been previously explored by \citet{WAL07} in N$_2$H$^{+}$ (1-0) at resolutions of 10 arcsec. The velocity information of the dense gas provided by these interferometric observations is restricted to the surroundings of the most massive cores and IRAS sources \citep[see Fig.~4 in][]{WAL07}. With only a factor of three difference in terms of resolution, the improved sensitivity and larger coverage of our millimeter line observations permit us to investigate the global properties of the dense gas velocity field within this region with unprecedented detail.

\begin{figure}
\centering
\includegraphics[width=0.9\linewidth]{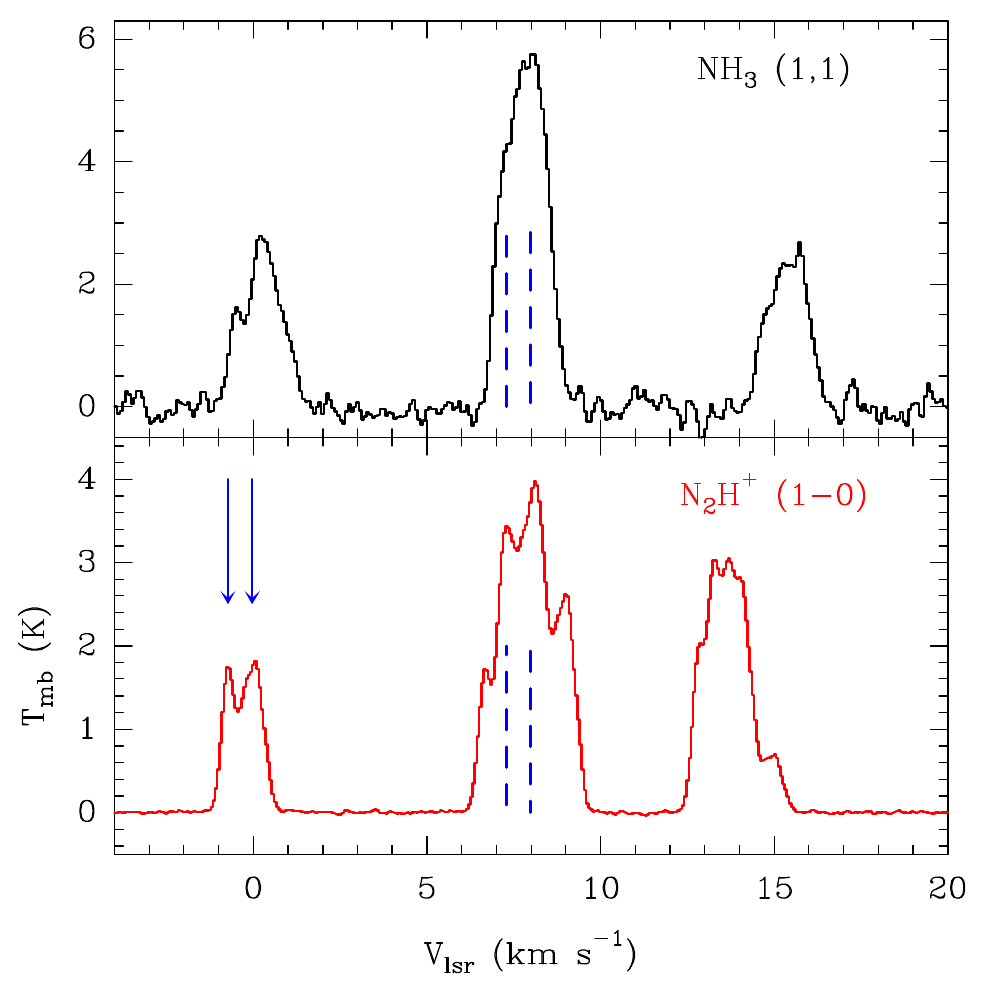}
\caption{Line blending in the NH$_3$ (1,1) (black; upper panel) and N$_2$H$^{+}$ (1-0) (red; lower panel) spectra observed at the central position (x,y)=(0,0) of our maps. We notice how the presence of 2 overlapping velocity components is clearly resolved out in the N$_2$H$^{+}$ isolated component (blue solid arrows) while it remains undistinguishable within the main hyperfine doublet of the NH$_3$ (1,1) line  (blue dashed lines).}
\label{fig:NGC1333_lineblending}
\end{figure}

The analysis in this section is carried out using the velocity information provided by our IRAM~30m N$_2$H$^{+}$ (1-0) spectra. Tracing the same gas component as the NH$_3$ (1,1) emission (Sect.~\ref{subsec:Nbearing}), these N$_2$H$^{+}$ observations are preferred due their better sensitivity, coverage, and resolution. Our choice is also justified by the observational advantages of the N$_2$H$^{+}$ emission in comparison to ammonia. First, the higher molecular weight of this species ($\mu$(N$_2$H$^{+}$)~=~29 vs $\mu$(NH$_3$)~=17 a.m.u.) translates into intrinsically narrower thermal lines. More importantly, the analysis of superposed velocity components is hampered by the intrinsic NH$_3$ (1,1) hyperfine line structure with 10 hyperfine lines, all found in doublets at close frequencies. On the contrary,  the isolated location and optically thin emission of the ($JF_1F=$~101-012) hyperfine component of N$_2$H$^{+}$, with $\tau (101-012)<$~0.9, allows a more precise decomposition of its emission in velocity.  While both tracers produce similar results in regions with a simple kinematic structure, the favourable emission properties of the N$_2$H$^{+}$ (1-0) spectra makes this line a more reliable tracer in the case of complex spectra with multiple components. An example of these advantages can be seen in Figure~\ref{fig:NGC1333_lineblending}.

\subsection{Line decomposition and  analysis}\label{subsec:kinematics:linedecomp}

  \begin{figure*}
   \centering
   \includegraphics[width=0.9\textwidth]{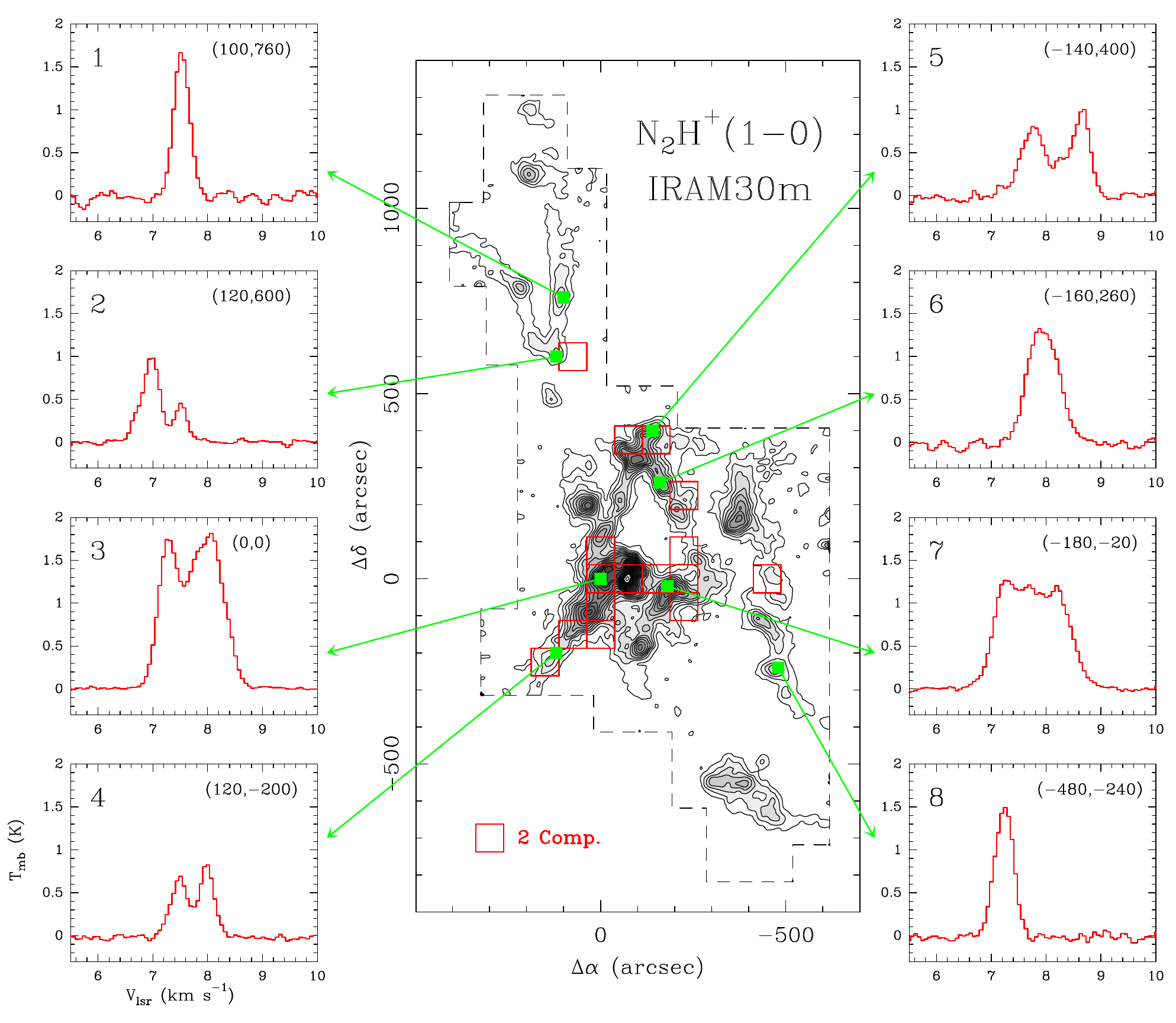}
      \caption{Dense gas kinematics in the NGC1333 proto-cluster traced by our IRAM~30m N$_2$H$^{+}$ observations. (Central panel) N$_2$H$^{+}$ integrated emission. The red squares enclose those subregions fitted with two independent components (see text), typically found surrounding the most prominent emission clumps detected in this molecule. 
      (Lateral subpanels) Representative N$_2$H$^{+}$ ($JF_1F=$~101-012) spectra (isolated component) within selected regions (spectra 1-8). The position of the spectra are indicated in the upper-left corner of each subpanel. We note the presence of two clearly separated components detected in N$_2$H$^{+}$ in spectra 2-5. Although still fitted with two components, a third line superposition is suggested by the shape of the line profile in spectrum 7. }
             \label{fig:NGC1333_spectra}
   \end{figure*}
   
  In Fig.~\ref{fig:NGC1333_spectra}, we present a representative subset of the N$_2$H$^{+}$ (1-0) spectra observed within the NGC1333 region (lateral subpanels 1-8). A direct inspection of the data reveals that approximately $\sim$~3/4 of the observed positions with detected N$_2$H$^{+}$ emission present a single-line, narrow line component (e.g., spectra 1, 6, and 8). In the other $\sim$~1/4 of our data, we find clear evidence of line multiplicity, typically in the form of a double-peaked profile (spectra 2-5). A similar behavior is observed in ammonia at those positions where the individual velocity components are well separated in velocity. Although not considered for our analysis, a single three-components spectrum is found coincident with the positions of the IRAS~2 source (spectrum 7). Remarkably, no significant differences are found in the observed linewidths of these individual N$_2$H$^{+}$ components along the cloud.

The widespread presence of double peaked spectra in the surroundings of the IRAS~2, 3, 4, and 6 sources has been reported by  \citet{WAL06} from the analysis of HCO$^+$ (1-0) and (3-2) lines. Walsh et al identified these complex HCO$^+$ spectra as asymmetric, self-absorbed profiles, generated by the relative motions between the dense gas and its infalling envelope. 
Sensitive to a denser gas component than the optically thick HCO$^+$ emission, the different velocity peaks detected in the isolated, optically thin N$_2$H$^{+}$ hyperfine component actually correspond to independent gas structures superposed along the line-of-sight. This conclusion is reinforced by the global distribution, continuity, and velocity coherence of these components at large scales (see also Sec.~\ref{subsec:kinematics:fibers}). Hereafter, our analysis assumes a direct correspondence between the individual dense gas structures within NGC1333 and the velocity components traced in our N$_2$H$^{+}$ spectra.

We have derived the kineamtic properties of each gas component detected in our N$_2$H$^{+}$ (1-0) spectra from the analysis of their individual hyperfine fits. 
The line decomposition method follows the fitting strategy introduced by \citet{HAC13} adapted to the improved resolution of our new IRAM~30m observations.
In short, the line fitting procedure is carried out after dividing the cloud in subfields of 5$\times$5 adjacent spectra, 75$\times$75 arcsec$^2$ each. For each of these groups, we obtain an average spectrum 
considering only those spectra with at least one emission channel with S/N~$\ge$~0.45~K (or 3$\sigma$(average) the average noise in our spectra) within the velocity range of the cloud. 
The fitting results of this local template (i.e., number of components, intensities, widths, and central velocities) are then used as initial guess values for the automatic fit of the spectra within the subgroup satisfying the above criteria. To ensure the quality of our analysis, only those spectra fitted with peak temperatures with S/N~$\ge$~3$\sigma$(spectrum) are considered for their analysis (see \citet{HAC13} for a detailed discussion of the fitting method). A similar procedure is applied in the fit of our NH$_3$ (1,1) and (2,2) observations, as well as our N$_2$H$^{+}$ data convolved to 60 arcsec (see Sect.~\ref{subsec:Nbearing}),  this time in subgroups 5$\times$5 spectra within a representative area of 150$\times$150 arcsec$^2$.

Among the total 5525 positions surveyed in N$_2$H$^{+}$ (1-0), 
we successfully obtain 2719 individual fits  ($\sim$~85\% in single peaked spectra, 15\% in double peaked), 1888 ($\sim$~70\%) of which have S/N~$\ge$~3. As indicated in Fig.~\ref{fig:NGC1333_spectra} (central panel), the detection of double peaked spectra is more common towards those regions with higher integrated N$_2$H$^{+}$ emission, like IRAS~3 or IRAS~4. Similarly, a total of 476, 140, and 595 high S/N components are recovered in the $\sim$~1000 spectra our NH$_3$ (1,1), NH$_3$ (2,2) (using gaussian fits), and N$_2$H$^{+}$ (1-0) maps with a resolution of 60 arcsec.
For each of the above components, our fits provide information about the line Full With Half Maximum (FWHM, $\Delta V$), the line centroid (V$_{lsr}$), and the peak line intensities (T$_{mb}$), either provided by direct measurements, in the case of the gaussian fits of the NH$_3$ (2,2) spectra, or determined in combination with the opacities ($\tau$) in the case of the N$_2$H$^{+}$ (1-0) and NH$_3$ (1,1) hyperfine fits.

\subsection{Gas velocity field: general properties}\label{subsec:kinematics:NTmotions}

We have investigated the main properties of the dense gas kinematics in the NGC1333 proto-cluster from the statistical analysis of the derived $\Delta V$ and V$_{lsr}$ values in our N$_2$H$^{+}$ lines. The independent study of these two observables allows us to infer the global characteristics of the gas motions both along the line-of-sight and across the cloud, respectively. Following standard practice, we have compared the magnitud of these motions with the expected (1D) H$_2$ sound speed c$_s$ (i.e., $\sigma_{Th}(H_2)=\sqrt{k_B T_K/\mu(H_2)}$~). For simplicity, all the gas traced in N$_2$H$^{+}$ is assumed to present an uniform T$_K=$~10~K, leading to a constant sound speed of c$_s$=$\sigma_{Th}(H_2)=$~0.19~km~s$^{-1}$
for an H$_2$ molecular gas in a mixture of 10\% He with a mean molecular mass of $\mu(H_2)=$~2.33.
With T$_K$ varying between 10-20~K within our maps (Sect.~\ref{subsec:gastemp}), this c$_s$ value is assumed to define a lower limit of the actual sound speed for the dense gas traced in N$_2$H$^{+}$.

To characterize the dense gas velocity field in NGC1333, we have derived the magnitude of the non-thermal motions along the line-of-sight at each of the positions with significant N$_2$H$^{+}$ emission. 
Its dispersion ($\sigma_{NT}$) is directly obtained from the observable $\Delta V$ after the subtraction of the thermal velocity dispersion of the gas ($\sigma_{Th}$) in quadrature:
\begin{equation}
	\sigma_{NT}=\left[\left(\frac{\Delta V}{\sqrt{8 log~2}}\right)^2-\frac{k_B T_K}{\mu(N_2H^+)} \right]^{1/2},
\end{equation}
where $\sigma_{Th}=\sqrt{k_B T_K/\mu(N_2H^+)}=$~0.053 km~s$^{-1}$ is the N$_2$H$^{+}$ thermal broadening at T$_K=$~10~K.

On average, the gas traced by this molecule presents an average value of $\left< \sigma_{NT}/c_s \right>=1.1\pm0.6$ with 1746 components (92.5\%) detected in N$_2$H$^{+}$ with S/N~$\ge$~3$\sigma$ presenting either subsonic (53.8\%; i.e., $\sigma_{NT}\le$~c$_s$) or transonic (38.7\%; i.e., ~c$_s < \sigma_{NT}\le$~2~c$_s$) velocity dispersions. Conversely, only 7.5\% of the N$_2$H$^{+}$ components exhibit supersonic non-thermal dispersions ($\sigma_{NT}>$~2~c$_s$). Most of the above transonic and supersonic components are, however, found in regions with higher gas temperatures in the surroundings of sources like IRAS 3 or IRAS 2 (Sect.~\ref{subsec:gastemp}). Their fractional contributions are therefore considered as upper limits for the actual dense gas velocity dispersion.

The above narrow, sonic-like velocity dispersions 
contrasts with the supersonic V$_{lsr}$ variations in NGC1333, with a maximum velocity difference of max(V$_{lsr}$)-min(V$_{lsr}$)~=~2.6~km~s$^{-1}$ (or >10~c$_{s}$). 
This property can be directly deduced from the presence of well-defined superposed components observed in our spectra (e.g., Fig.~\ref{fig:NGC1333_spectra}). 
We have quantified these gas velocity differences from the dispersion of the observed V$_{lsr}$ resulting into values of $\left<\sigma\mathrm{V}_{lsr}/ \mathrm{c}_s\right>$=2.6.
With $\{\sigma_{NT}, \sigma V_{lsr} \}=\{ 1.1, 2.6 \}\times$~c$_s$, our results indicate that the largest velocity differences inside the NGC1333 proto-cluster are generated by the relative motions between individual gas components at large scales (aka macroscopic motions) rather than their local dispersions (i.e., microscopic motions) \citep[see ][ for a discussion]{HAC16b}.

\subsection{Quiescent fibers in NGC1333}\label{subsec:kinematics:fibers}

   \begin{sidewaystable*} 
\caption{Fibers in NGC1333: physical properties} \label{table:NGC1333_fiberprop} 
\centering 
\begin{tabular}{l c c c c c c c c c c c c c c c} 
\hline 
\hline 
ID &	$ \alpha_0$ &	$ \delta_0$ &	L &	R$_{max}$ &	AR &	M &	m$_{lin}$ &	m$_{lin}$/m$_{crit}$ &	$\left< V_{lsr} \right>$ &	$\left< \sigma V_{lsr} \right>$ &	$\left< \sigma_{NT}/c_s \right>$ &	$\left< \nabla V_{lsr}|_{global} \right>$ &	$\left< \nabla V_{lsr}|_{local} \right>$ &	N (cores) & $\left< D(cores) \right>$\\ 

&	('') &	('') &	(pc)&	(pc)&	&	(M$_{\sun}$) &	(M$_{\sun}$~pc$^{-1}$) &	&	(km s$^{-1}$) &	(km s$^{-1}$) &	 &	(km s$^{-1}$~pc$^{-1}$) &	(km s$^{-1}$~pc$^{-1}$) &	& (pc) \\  
(1) & (2) & (3) & (4) & (5) & (6) & (7) & (8) & (9) & (10) & (11) & (12) & (13) & (14) & (15) & (16) \\
\hline 
    1 &	 -490 &	 -238 &	0.48 &	0.05 &	8.8 &	10.1 &	22.0 &	0.85 &	7.3 &	0.07 &	0.76 &	0.4 &	0.8 &	4 & 0.085\\ 
    2 &	 -488 &	  -45 &	0.16 &	0.05 &	3.2 &	2.7 &	16.7 &	0.63 &	7.6 &	0.09 &	0.80 &	0.4 &	2.1 &	2 & 0.049\\ 
    3 &	 -416 &	 -584 &	0.40 &	0.10 &	3.8 &	16.4 &	41.4 &	1.50 &	6.9 &	0.29 &	0.83 &	2.5 &	2.6 &	4 & 0.077\\ 
    4 &	 -401 &	  162 &	0.30 &	0.07 &	4.3 &	15.4 &	51.0 &	1.03 &	8.1 &	0.13 &	1.42 &	0.8 &	1.8 &	3 & 0.052\\ 
    5 &	 -254 &	   -6 &	0.37 &	0.08 &	4.8 &	13.8 &	37.1 &	0.83 &	7.9 &	0.21 &	1.34 &	1.7 &	2.5 &	4 & 0.059\\ 
    6 &	 -219 &	  385 &	0.21 &	0.08 &	2.8 &	2.9 &	14.7 &	0.40 &	8.4 &	0.12 &	1.10 &	1.4 &	2.3 &	2 & 0.063\\ 
    7 &	 -145 &	  286 &	0.35 &	0.11 &	3.2 &	22.6 &	69.9 &	1.40 &	7.8 &	0.12 &	1.41 &	0.8 &	1.3 &	5 & 0.071\\ 
    8 &	 -122 &	 -189 &	0.48 &	0.10 &	4.6 &	15.7 &	32.9 &	0.89 &	7.2 &	0.11 &	1.12 &	0.2 &	1.5 &	2 & 0.134\\ 
    9 &	  -16 &	  136 &	0.47 &	0.09 &	5.0 &	28.1 &	58.5 &	1.30 &	8.4 &	0.17 &	1.30 &	0.7 &	1.4 &	5 & 0.075\\ 
   10 &	   18 &	  -55 &	0.34 &	0.11 &	3.0 &	28.1 &	82.5 &	1.36 &	7.8 &	0.12 &	1.62 &	0.7 &	1.5 &	4 & 0.058\\ 
   11 &	  131 &	 -227 &	0.25 &	0.03 &	9.7 &	3.1 &	12.3 &	0.36 &	7.6 &	0.17 &	1.01 &	2.4 &	2.5 &	4 & 0.046\\ 
   12 &	  127 &	  917 &	1.01 &	0.09 &	11.1 &	20.1 &	20.7 &	0.91 &	7.4 &	0.10 &	0.62 &	0.2 &	0.9 &	4 & 0.199\\ 
   13 &	  172 &	 1252 &	0.18 &	0.04 &	4.9 &	2.2 &	11.7 &	0.53 &	7.2 &	0.05 &	0.58 &	0.1 &	1.0 &	1 & ---\\ 
   14 &	  356 &	  914 &	0.23 &	0.04 &	5.9 &	2.8 &	11.9 &	0.52 &	7.8 &	0.06 &	0.63 &	0.4 &	1.3 &	1 & ---\\ 
\hline 
Mean &	 --- &	 --- &	0.37 &	0.07 &	5.4 &	13.2 &	34.5 &	0.89 &	7.7 &	0.13 &	1.04 &	0.9 &	1.7 &	3.75 & 0.081 \\ 
\hline 
\end{tabular} 
\tablefoot{Columns: (1) Fiber ID; (2) \& (3) central position in radio offsets respect to the map center; (4) fiber length, L; (5) maximum radius, R$_{max}$; (6) Aspect ratio, $AR=L/R_{max}$; (7) total mass, M$_{tot}$; (8) mass-per-unit-length, M$_{lin}$; (9) virial mass, M$_{vir}$; (10) mean gas velocity, $\left< V_{lsr} \right>$; (11) mean line centroid dispersion, $\left< \sigma V_{lsr} \right>$; (12) mean non-thermal velocity dispersion, $\left< \sigma_{NT} \right>$; (13) mean, global velocity gradient, $\left<\nabla V_{lsr}|_{global} \right>$; (14) mean local gradient, $\left<\nabla V_{lsr}|_{local} \right>$; (15) number of embedded core candidates; (16) mean nearest distance between core candidates along the fiber, $\left< D(cores) \right>$ (see text for details). Mean values for all the above parameters are indicated in the last row.
} 
\end{sidewaystable*} 

\begin{figure*}
	\centering
	\includegraphics[width=0.5\linewidth]{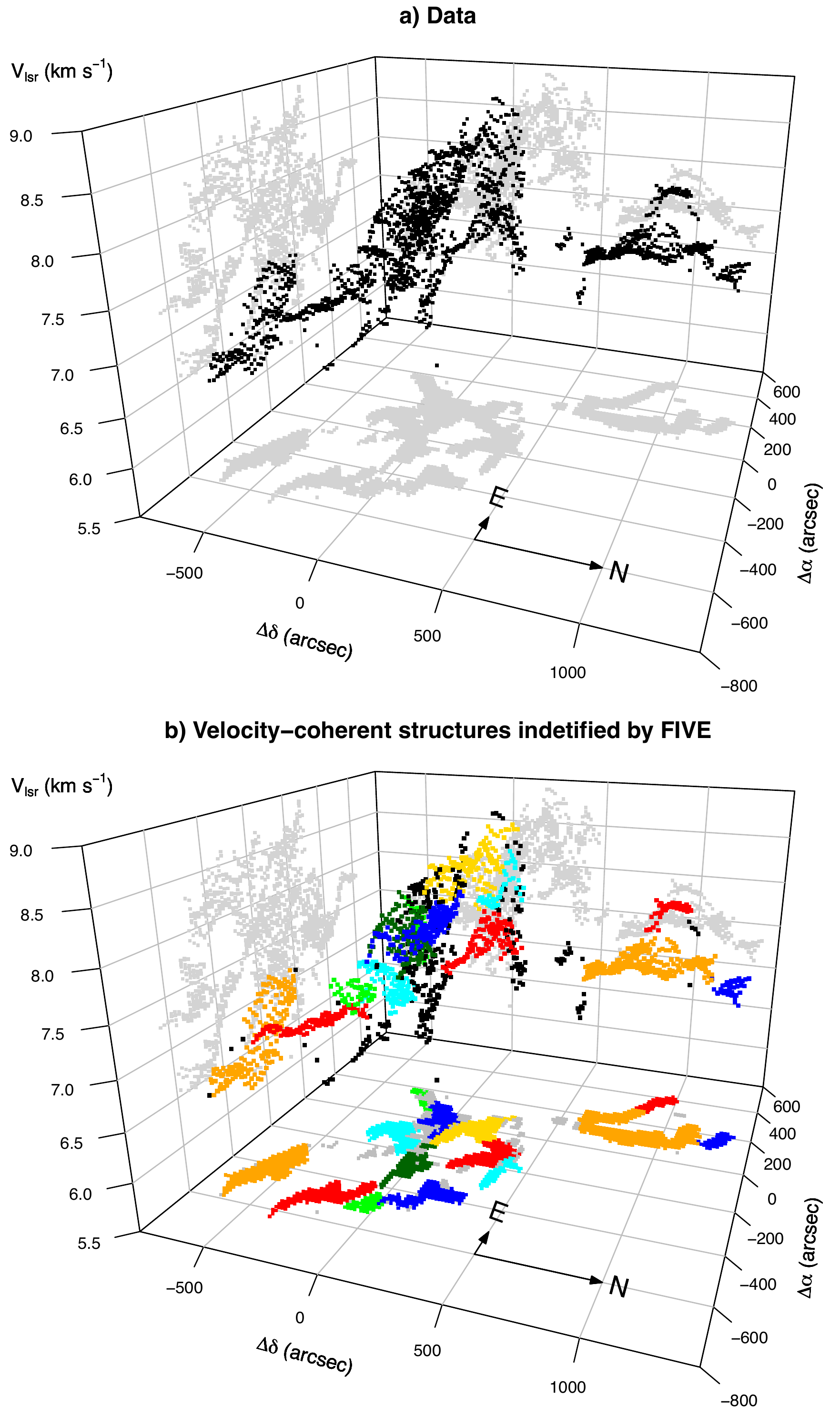}
	\includegraphics[width=0.40\linewidth]{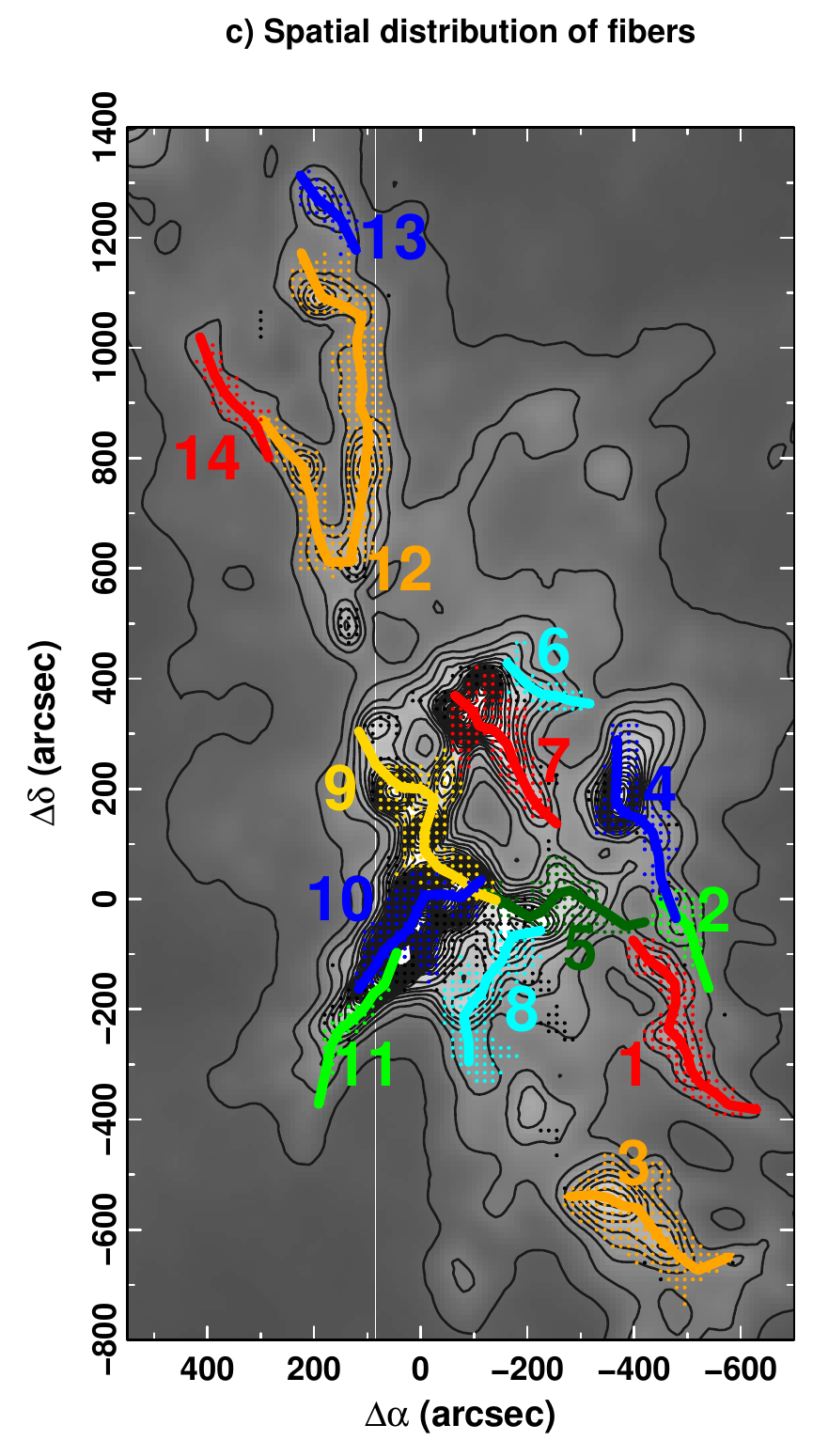}
	\caption{Identification of velocity-coherent structures (aka fibers) within NGC1333 in both Position-Position-Velocity (PPV) space (a and b subpanels) and Position-Position (PP) space (c subpanel). 
		The original data points are indicated by black dots. Their projection in the corresponding sub-space (either PP or PV) is indicated in grey in subpanels a and b. 
		The corresponding structures identified by FIVE is colour coded in subpanels b and c. The main axis of each of the 14 structures found in NGC1333 is represented by solid thick lines in panel c superposed to Herschel-Planck total column density map of \citet{ZAR16} (grey scale; contours equally spaced every A$_V=$~5$^{mag}$). We notice that the orientation of the PPV diagrams (a and b subpanels) is rotated 90~deg respect to the PP maps (c subpanel). 
	}
	\label{fig:NGC1333_PPV}
\end{figure*}

In addition to the small variations found in the gas velocity dispersion, the analysis of the line velocity centroids reveals a high degree of kinematic organization within NGC1333.
Figure~\ref{fig:NGC1333_PPV}a shows the spatial distribution of the line centroids in the 3D Position-Position-Velocity (PPV) space.
 Different groups of points are easily recognized by a characteristic spatial and velocity continuity in scales on the order of 0.3-0.5~pc. 
 Several of these structures seem to converge towards the centre of this cluster coincident with those positions with observed spectra with multiple components. A similar kinematic complexity and spectral multiplicity has been previously identified with the line-of-sight superposition of structures in the so-called bundles of fibers, first observed in the B213-L1495 filament in Taurus \citep{HAC13}. Our new observations indicate that the presence of intricate fiber networks might not be restricted to isolated filaments, but could also explain the internal gas substructure of more complex regions like NGC1333.

We have investigated the internal gas structure within NGC1333 using an improved version of the Friends-In-VElocity (FIVE; version v.1.1) analysis technique \citep{HAC13}. In a nutshell, FIVE is specifically designed to identify and reconstruct continuous, velocity-coherent structures from the progressive association of different gas components in the PPV space. A full description of this algorithm and its performance is described in \citet{HAC13}. Compared to the original selection proposed for this algorithm based on a S/N cut (v.1.0), the improved version (v.1.1) uses an absolute peak intensity threshold (I$_0$) for the first selection of points defining the spine of the velocity coherent structures (see steps 1 \& 2 in \citet{HAC13}). This change replaces the previous sensitivity dependent criterion by a physically motivated selection threshold, allowing direct comparisons with additional datasets. The rest of the algorithm remains unaltered. For the analysis presented in this work, we have adopted
values of I$_0$=~0.5~K, L$_{friends}=$~60~arcsec (or 2 beams), and $\nabla V_{lsr}=$~4~km~s$^{-1}$~pc$^{-1}$
\citep[see a full description of each of these parameters in ][]{HAC13}. The selected algorithm parameters guarantee that the maximum velocity difference between points connected at a distance L$_{friends}$ is smaller than the mean observed $\left< \Delta V(N_2H^{+}) \right>$, that is, L$_{friends}\times \nabla V_{lsr} < 0.49$~km~s$^{-1}$. Also, it allows a direct comparison with the previous results obtained by \citet{HAC13} in Taurus using similar linking properties, that is (L$_{friends}\times\nabla V_{lsr})|_{B213}\simeq$~(L$_{friends}\times\nabla V_{lsr})|_{NGC1333}$, in both B213-L1495 and NGC1333 analysis.

Using the above FIVE parameters, we recover a total of 14 independent velocity-coherent structures in NGC1333. Each of these structures is  identified in the PPV space diagram presented in Fig.~\ref{fig:NGC1333_PPV}b \footnote{For visualization purposes, an animation showing an artistic rendering of the NGC1333 fibers in the PPV space can be found at the following link: http://home.strw.leidenuniv.nl/$\sim$hacar/NGC1333\_PPV\_movie.gif}. Their spatial distribution in the Position-Position space (aka plane-of-the-sky) is also shown in Fig.~\ref{fig:NGC1333_PPV}c superposed to the Herschel column density maps obtained by \citet{ZAR16}. 
Among the 1888 components with S/N~$\ge$~3 analysed in this cloud (black squares), 1555 of them ($\sim$~83\%) are associated to velocity-coherent structures (coloured points). Two additional velocity-coherent structures can be recognized in the vicinity of IRAS~3 and IRAS~2 protostars (see black dots in our PPV plot), but their identification is hampered by the lower number of points defining their axis and the larger local velocity gradients (steps 1 and 2 of FIVE). Additional tests including larger velocity gradients (i.e., $\nabla V_{lsr}$~=5-6~km~s$^{-1}$~pc$^{-1}$) allow us to recover a larger fraction of points ($>$~90\%), including the above two additional structures. As a drawback, the use of these gradients also produces the artificial merging of clearly separated structures in complex environments, like the surroundings of IRAS~3 and IRAS~6 sources (e.g., velocity-coherent structures 7 and 9). In contrast, lower $\nabla V_{lsr}$ values tend to artificially fragment several of the well behaved substructures with continuous but larger internal gradients (e.g., structure 3). Although the final identification and association of these velocity-coherent structures depend on the fine tuning of the algorithm parameters, most of their mean properties remain unaltered for $\nabla V_{lsr}$ values between  $\sim$~3-6~km~s$^{-1}$~pc$^{-1}$.

\begin{figure*}
   \centering
   \includegraphics[width=1.0\linewidth]{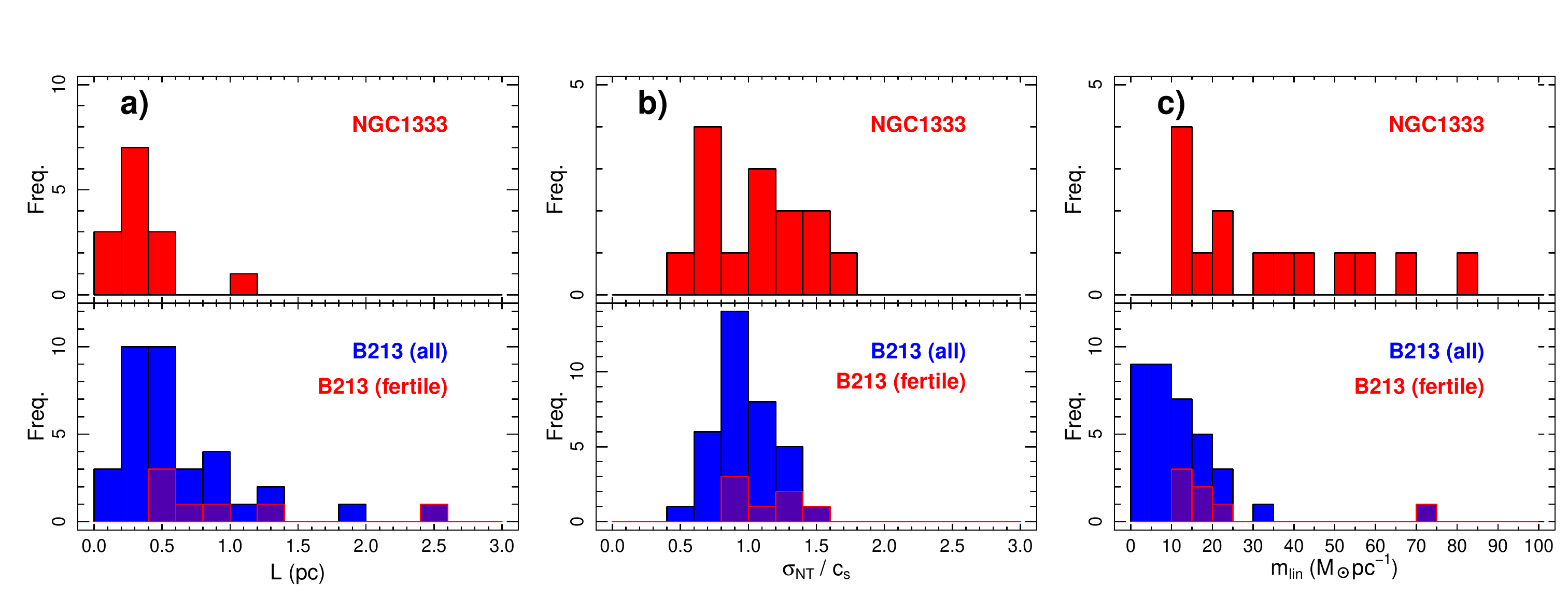}
      \caption{Comparison of the fiber properties in NGC1333 (red solid histograms; upper panel) and B213-L1495 \citep[blue solid histograms, lower panel;][]{HAC13}. The main properties of the subgroup of fertile fibers found in B213-L1495 are highlighted in red in the lower panel.
      From left to right: (a) total fiber length; (b) internal non-thermal motions (in units of the sound speed c$_s$); and (c) mass-per-unit-length.
      }
             \label{fig:NGC1333_histos}
   \end{figure*}
   
The main physical properties of the 14 velocity-coherent structures identified within NGC1333 are summarized in Table~\ref{table:NGC1333_fiberprop}. These parameters are derived following the methods and analysis introduced by \citet{HAC13}.
 Although not required by FIVE, we notice that most of the structures recovered there present elongated and filamentary shapes. We have defined the main axis of each velocity-coherent structure using the 2-steps approximation introduced in the Appendix~B of \citet{HAC13} (see also Fig.~\ref{fig:NGC1333_PPV}c).
For each of the structures recovered by FIVE (column 1),
Table~\ref{table:NGC1333_fiberprop} includes values for their central position (x$_0$, y$_0$) (cols. 2 \& 3), length along its main axis (L; col. 4), maximum radius (R$_{max}$; col. 5), and aspect ratio ($AR=L/R_{max}$; col. 6). In each case, the total mass (M; col. 7) is derived from the total N$_2$H$^+$  integrated emission converted into a total column density using Eq.~\ref{eq:N2Hp_H2}. From it, the mass-per-unit-length (m$_{lin}$; col. 8) is simply calculated from the ratio M/L. This m$_{lin}$ is also compared to the corresponding critical mass m$_{crit}=\frac{2 \sigma_{eff}^2}{G}$ (col. 9), including both thermal (c$_s$) and non-thermal motions ($\sigma_{NT}$) with $\sigma_{eff}^2=c_s^2+\sigma_{NT}^2$ (see Sect.~\ref{sec:fertile_fibers}). Describing their internal kinematics, we obtain their mean gas velocity ($\left< V_{lsr} \right>$; col. 10) and line centroid dispersion ($\left< \sigma V_{lsr} /c_s \right>$; col. 11), as well as their mean non-thermal velocity dispersion along the line-of-sight ($\left< \sigma_{NT} \right>$; col. 12), the last two in units of the sound speed c$_s$. Additionally, we report values for both their mean velocity gradient ($\left<\nabla V_{lsr}|_{global} \right>$; col. 13), calculated from the global linear fit of the gas velocities along the entire fiber, as well as their
mean local gradient ($\left<\nabla V_{lsr}|_{local} \right>$; col. 14) estimated in intervals of 0.05~pc \citep[see][for details]{HAC13}.
 Table~\ref{table:NGC1333_fiberprop} also includes the number of cores embedded in each velocity-coherent structure (col. 15) identified in Sect.~\ref{sec:fragmentation}. Mean values of each of the above properties are indicated in the last row of this table.

The physical characteristics of the 14 structures identified in the NGC1333 proto-cluster resemble the physical properties defining the so-called velocity-coherent fibers previously identified in Taurus by \citet{HAC11} and \citet{HAC13}. An aspect ratio (AR) of $\sim$~5 indicates that the structures are filamentary. In addition, the gas presents transonic internal motions, both $\left< \sigma_{NT} \right>$ and $\left< \sigma V_{lsr} /c_s \right>$ of the order of the sound speed c$_s$,  along their average total length of L$\sim$~0.4~pc. Moreover, their internal velocity field is characterized by smooth internal velocity gradients with both $\left<\nabla V_{lsr}|_{local} \right>$ and $\left<\nabla V_{lsr}|_{global} \right>$ on the order of $\sim$~1-2~km s$^{-1}$~pc$^{-1}$. 
As highlighted in the different histograms presented in Figs.~\ref{fig:NGC1333_histos}a and \ref{fig:NGC1333_histos}b, the values derived for the different velocity-coherent structures in NGC1333 (red histograms, upper panels) are in close agreement with the range of values defining the fiber properties in B213-L1495 \citep{HAC13} (blue histograms, lower panels). Similar results can be found from the comparisons of their typical gradients and internal motions (not shown here). Based on their analogue kinematic and morphological properties, we identify the above 14 structures as velocity-coherent fibers. 

Figure~\ref{fig:NGC1333_PPV}c shows the direct correlation of the detected fibers in NGC1333 with mass distribution traced in the previous Herschel maps. Despite being independently obtained from our analysis of the N$_2$H$^{+}$ kinematics, we find an excellent agreement between the fiber axis defined by FIVE (colored segments) and the crest and high column density features detected in the dust continuum maps (contours in the background image). In absolute term, the total mass in fibers corresponds to $\sim$~75\% of the total mass of dense gas detected in our maps (Sect.~\ref{DGMF}). Our results demonstrate that most of the apparent kinematic and structural complexity of this young proto-cluster is actually originated by its underlying network of sonic-like fibers.

\subsection{Non-thermal motions and stability of massive fibers}\label{sec:fertile_fibers}

While sharing the same average kinematic properties, the mass-per-unit-length (m$_{lin}$) of the NGC1333 fibers extends towards the high-mass end range of masses reported in the case of Taurus. This difference can be easily appreciated in the Filament Linear Mass Function (FLMF) of these two regions shown in Fig.~\ref{fig:NGC1333_histos}c. Based on the N$_2$H$^{+}$ detection of dense cores within the Taurus fibers, \citet{HAC13} and \citet{TAF15} have distinguished between two distinct fiber families: those with embedded N$_2$H$^{+}$ dense cores are referred to as {\it fertile} while those devoid of them and without significant detection of dense gas are classified as {\it sterile}. Statistically speaking,  fertile fibers have m$_{lin}\gtrsim$~15~M$_{\sun}$~pc$^{-1}$. On the contrary, typical values of  m$_{lin}\lesssim$~15~M$_{\sun}$~pc$^{-1}$ are found for their sterile counterparts. In comparison, all the NGC1333 fibers exhibit a high degree of fragmentation with multiple embedded cores per fiber (see also Sect.~\ref{sec:fragmentation}) with m$_{lin}$ values ranging between $\sim$~12-90~M$_{\sun}$~pc$^{-1}$. Their similarities with the more massive Taurus fibers with embedded cores suggest that the 14 fibers found in NGC1333 correspond to the more active and star-forming, fertile type (see red open histograms in the lower panel). 

The above observational selection effect can be explained by the use of N$_2$H$^{+}$ as line tracer for the study of the gas kinematics in NGC1333.  In the case of B213,  $\sim$~1/4 of the fibers are identified as fertile, while the other $\sim$~3/4 are considered to be sterile. This ratio is consistent with the reported FLMF in this cloud, approximately with $\Delta N/\Delta m_{lin}\propto \Delta m_{lin}^{-0.3}$, for the mass range between m$_{lin}=$~[10-30]~M$_{\sun}$~pc$^{-1}$ and a effective mass threshold of $\sim$~20~M$_{\sun}$~pc$^{-1}$. 
Following the same distribution, a total of $\sim$~50 sterile fibers would be expected along the entire NGC~1333 ridge. 
Confirming this hyphotesis would require a detailed analysis of tracer at lower densities (e.g., C$^{18}$O).  Nevertheless, several additional filamentary structures can be already recognized in the column density maps detected by Herschel (e.g., see position (x,y)=(-200,800) Fig.~\ref{fig:NGC1333_largescale}).

\begin{figure}
   \centering
   \includegraphics[width=0.95\linewidth]{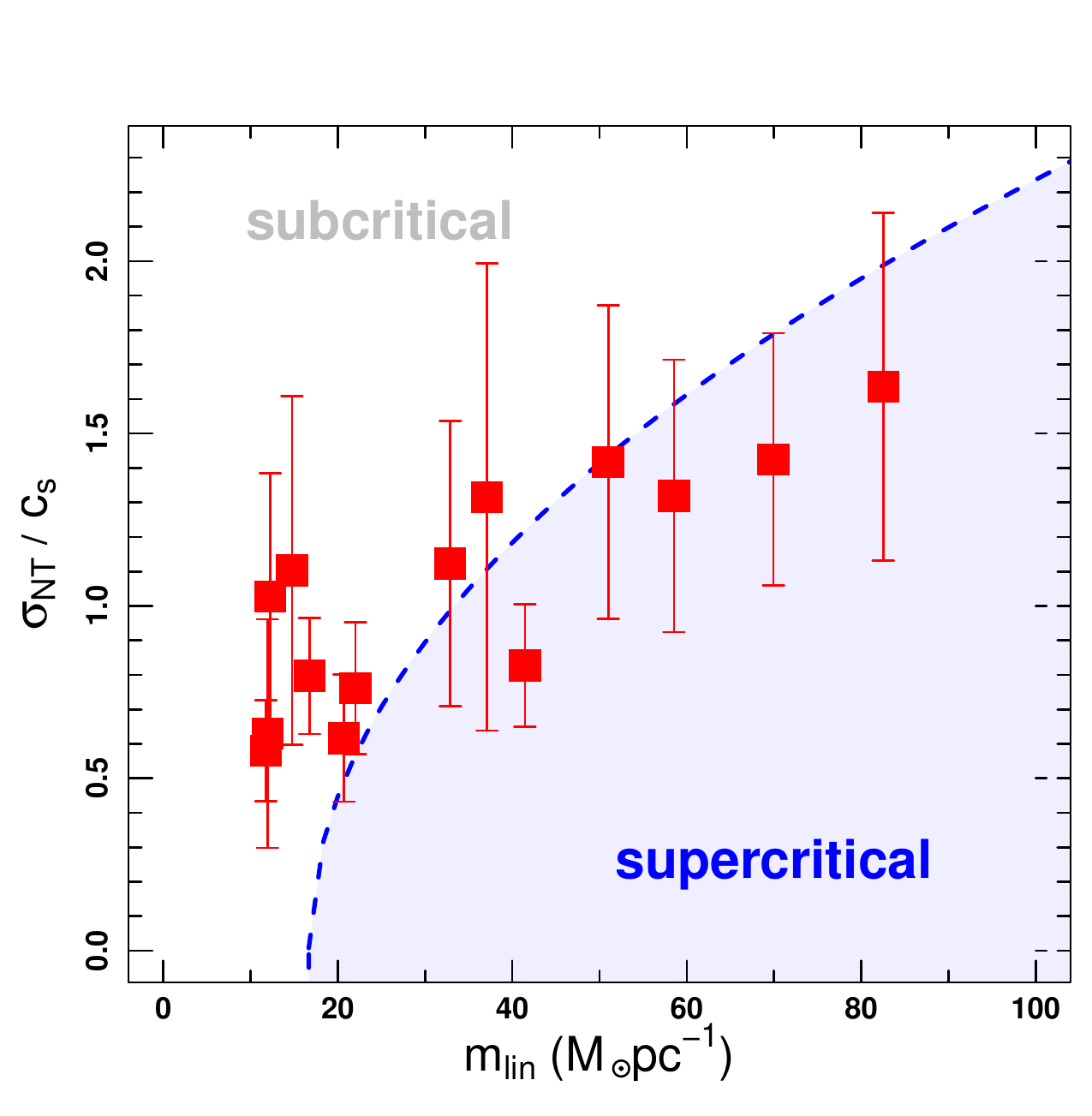}
      \caption{Correlation between non-thermal motions ($\sigma_{NT}$) and the mass-per-unit-length (m$_{lin}$) in the NGC1333 fibers (red squares). The blue dashed line denotes the critical m$_{lin}$ for a filament in hydrostatic equilibrium, m$_{crit}(\sigma_{eff})=\frac{2\sigma_{eff}^2}{G}$, with $\sigma_{eff}^2=c_s^2+\sigma_{NT}^2$, separating the subcritical (i.e., m$_{lin}\le$~m$_{crit}$ and supercritical regimes (i.e., m$_{lin}>$~m$_{crit}$).
      The vertical error bars correspond to the 1$\sigma$ variations for the observed $\sigma_{NT}$ in each fiber.
      }
             \label{fig:NGC1333_virial}
   \end{figure}
   
   The stability of filamentary structures is commonly evaluated from the comparison of their observed mass per-unit-length m$_{lin}$ with the expected critical mass of an infinite filament in hydrostatic equilibrium, m$_{crit}(\sigma_{eff})=\frac{2\sigma_{eff}^2}{G}$ \citep{STO63,OST64}. In an extended practice, $\sigma_{eff}$ is assumed as the isothermal sound speed c$_{s}$(T$_K$) (see Sect.~\ref{subsec:kinematics:NTmotions}) leading to a  values of m$_{crit}(\sigma_{eff}=\mathrm{c}_{s})=$~16.6~M$_\sun$~pc$^{-1}$ at T$_K$=~10~K.  In the framework of this hydrostatic filament, subcritical filaments (i.e., m$_{lin}\le$~m$_{crit}$) are stabilized by their own internal pressure while supercritical filaments (i.e., m$_{lin}>$~m$_{crit}$) are expected to collapse into a spindle under their own weight in timescales comparable to their free-fall time $\tau_{ff}$ \citep{INU97}. Under this premise, an apparent contradiction arises on the interpretation of our results in NGC1333. How can the above 14 fertile fibers identified within this region, most of them m$_{lin}>$~16.6~M$_\sun$~pc$^{-1}$, survive during timescales comparable to $\tau_{dense} \sim 5\ \tau_{ff}$ (see Sect.~\ref{subsec:dense_proto})?
   
   An additional contribution to the filament support is provided by the non-thermal motions $\sigma_{NT}$. In the case of isotropic motions (i.e., microscopic turbulence), their contribution acts as internal pressure source increasing the effective $\sigma_{eff}^2=c_s^2+\sigma_{NT}^2$ and, therefore, their critical hydrostatic mass m$_{crit}$. Filaments with linear masses m$_{lin}>16.6$~M$_\sun$~pc$^{-1}$ are not necessary unstable unless m$_{lin}$ is larger than the corresponding m$_{crit}(\sigma_{eff})$. 

In Fig.~\ref{fig:NGC1333_virial}, we explore the correlation between the observed m$_{lin}$ and $\sigma_{NT}$ values in the 14 fibers identified in NGC1333 (red squares). For all these objects, we find a monotonic increase of the internal non-thermal motions with m$_{lin}$.
Superposed to our observational data, the blue dashed line indicates the value of m$_{crit}$ as a function of $\sigma_{NT}$, assuming a thermal contribution equal to c$_s$. At a given $\sigma_{NT}$ value, this line separates the two sub- and supercritical regimes.
The close correspondence between the derived m$_{crit}$ values with the observed m$_{lin}$ measurements (see also col. 9 in Table~\ref{table:NGC1333_fiberprop}) suggests that, despite their large m$_{lin}$, the observed fibers in NGC1333 remain close to equilibrium. 

A similar correlation between m$_{lin}$ and $\sigma_{NT}$ has been proposed by \citet{ARZ13} from the analysis of filaments in both local clouds like Aquila, Polaris, or IC~5146, and massive filaments like the DR21 ridge. The increase of the internal velocity dispersion in supercritical filaments is interpreted by these authors as the result of the contraction and/or infall within these objects. The confirmed presence of multiple line components and superposed structures in several of these filaments not resolved by Arzoumanian et al (e.g., \citet{FER14}), complicates the comparison with the individual and velocity-coherent fibers in this work. 
Still, the observed correlation between the fiber line masses and their observed non-thermal motions indicates a likely connection with the accretion process in these objects.
Independently of its origin, the observed increase of the internal motions with m$_{lin}$ provide additional support against gravity, explaining the long lifetimes of the NGC1333 fibers. 

\subsection{Fiber fragmentation and the origin of cores in NGC1333}\label{sec:fragmentation}

\begin{table*} 
\label{table:NGC1333_coreprop} 
\caption{Core candidates in NGC1333: derived properties} 
\centering 
\begin{tabular}{l c c c c c c c c c c c} 
\hline 
\hline 
Core &	Fiber &	$ \Delta\alpha$ &	$ \Delta\delta$ &	I$_{peak}$ &	I$_{peak}$/$\mathrm{I}_0$  &	FWHM &	M &	V$_{lsr}$ &	$\sigma_{NT}/c_s$ &	Ros08 &	 T$_K$ \\ 
&	&	('') &	('') &	(K) &	 &	(pc)&	(M$_{\sun}$) &	(km s$^{-1}$) &	 & (ID) &	(K) \\  
(1) &	(2) &	(3) &	(4) &	(5) &	(6) &	(7) &	(8) &	(9) &	(10) &	(11) &	(12) \\  
\hline 
    1 &	    1 &	 -511 &	 -315 &	0.6 &	1.5 &	0.099 &	0.38 &	7.19 &	0.5 &	  --- &	--- 	 \\ 
    2 &	    2 &	 -489 &	  -58 &	0.8 &	1.5 &	0.087 &	0.37 &	7.75 &	0.8 &	  --- &	--- 	 \\ 
    3 &	    2 &	 -480 &	  -16 &	0.6 &	1.4 &	0.120 &	0.52 &	7.70 &	0.8 &	  --- &	--- 	 \\ 
    4 &	    1 &	 -475 &	 -167 &	1.5 &	1.9 &	0.080 &	0.72 &	7.34 &	0.8 &	  --- &	--- 	 \\ 
    5 &	    1 &	 -475 &	 -242 &	2.1 &	2.2 &	0.072 &	0.91 &	7.27 &	0.7 &	   40 &	10.8 	 \\ 
    6 &	    3 &	 -463 &	 -630 &	1.0 &	1.4 &	0.089 &	0.65 &	6.64 &	0.8 &	   41 &	10.6 	 \\ 
    7 &	    3 &	 -439 &	 -505 &	0.7 &	1.5 &	0.064 &	0.22 &	6.74 &	1.0 &	   43 &	10.5 	 \\ 
    8 &	    4 &	 -432 &	  127 &	2.0 &	1.9 &	0.081 &	0.91 &	7.92 &	1.9 &	  --- &	--- 	 \\ 
    9 &	    1 &	 -429 &	 -113 &	2.4 &	2.1 &	0.064 &	0.71 &	7.42 &	1.0 &	   44 &	12.4 	 \\ 
   10 &	    3 &	 -379 &	 -553 &	1.8 &	1.6 &	0.087 &	1.02 &	7.01 &	0.9 &	   46 &	10.5 	 \\ 
   11 &	    4 &	 -378 &	  198 &	3.0 &	1.6 &	0.078 &	1.26 &	8.17 &	1.4 &	   47 &	11.7 	 \\ 
   12 &	    4 &	 -371 &	  165 &	3.4 &	1.7 &	0.072 &	1.40 &	7.98 &	1.4 &	   48 &	11.7 	 \\ 
   13 &	    3 &	 -340 &	 -541 &	1.6 &	1.5 &	0.087 &	0.93 &	7.36 &	0.8 &	   50 &	10.5 	 \\ 
   14 &	    5 &	 -292 &	    4 &	1.4 &	1.5 &	0.101 &	1.05 &	8.26 &	1.1 &	   51 &	10.8 	 \\ 
   15 &	    5 &	 -277 &	  -37 &	2.3 &	1.7 &	0.076 &	1.08 &	7.97 &	1.3 &	  --- &	--- 	 \\ 
   16 &	    5 &	 -273 &	   40 &	1.8 &	2.0 &	0.077 &	0.85 &	8.25 &	1.0 &	   52 &	11.3 	 \\ 
   17 &	    6 &	 -272 &	  358 &	0.7 &	1.5 &	0.074 &	0.28 &	8.35 &	0.8 &	  --- &	--- 	 \\ 
   18 &	    6 &	 -220 &	  376 &	0.8 &	1.4 &	0.091 &	0.40 &	8.31 &	0.9 &	  --- &	--- 	 \\ 
   19 &	    7 &	 -207 &	  180 &	1.9 &	1.8 &	0.067 &	0.65 &	7.64 &	1.1 &	   56 &	13.8 	 \\ 
   20 &	    5 &	 -197 &	  -30 &	6.6 &	3.1 &	0.074 &	2.67 &	7.74 &	2.1 &	   58 &	16.5 	 \\ 
   21 &	    7 &	 -169 &	  251 &	3.5 &	2.2 &	0.086 &	1.56 &	7.77 &	1.3 &	   59 &	14.5 	 \\ 
   22 &	    7 &	 -163 &	  307 &	2.4 &	1.6 &	0.072 &	0.99 &	7.74 &	1.7 &	   57 &	16.4 	 \\ 
   23 &	    8 &	 -154 &	  -82 &	3.7 &	2.1 &	0.066 &	1.24 &	7.20 &	1.1 &	  --- &	--- 	 \\ 
   24 &	    7 &	 -125 &	  377 &	3.5 &	2.0 &	0.082 &	1.67 &	7.81 &	1.6 &	   64 &	14.4 	 \\ 
   25 &	    8 &	  -99 &	 -185 &	3.6 &	2.0 &	0.071 &	1.50 &	7.08 &	1.4 &	   65 &	12.5 	 \\ 
   26 &	    7 &	  -98 &	  322 &	3.8 &	1.7 &	0.068 &	1.46 &	7.92 &	1.5 &	   66 &	16.4 	 \\ 
   27 &	  --- &	  -74 &	   -9 &	9.1 &	2.8 &	0.088 &	5.04 &	7.39 &	2.1 &	   68 &	16.4 	 \\ 
   28 &	    9 &	  -73 &	   40 &	7.6 &	2.7 &	0.062 &	2.27 &	8.41 &	1.4 &	   67 &	16.3 	 \\ 
   29 &	   10 &	  -67 &	    7 &	6.9 &	1.9 &	0.068 &	2.07 &	7.72 &	1.9 &	   68 &	16.4 	 \\ 
   30 &	    9 &	  -55 &	  211 &	2.0 &	1.8 &	0.084 &	1.00 &	8.47 &	1.3 &	   69 &	13.6 	 \\ 
   31 &	    9 &	  -33 &	   44 &	7.5 &	2.7 &	0.083 &	2.90 &	8.48 &	1.4 &	   70 &	14.8 	 \\ 
   32 &	   10 &	  -28 &	    9 &	5.4 &	1.6 &	0.075 &	2.00 &	7.87 &	2.0 &	   70 &	14.8 	 \\ 
   33 &	    9 &	   -8 &	  129 &	2.6 &	1.8 &	0.076 &	1.13 &	8.53 &	1.2 &	   72 &	12.6 	 \\ 
   34 &	   10 &	   -1 &	   -2 &	3.7 &	1.6 &	0.075 &	1.74 &	7.72 &	1.8 &	   73 &	12.3 	 \\ 
   35 &	  --- &	   23 &	  -91 &	5.1 &	2.4 &	0.069 &	1.95 &	7.29 &	1.9 &	   75 &	 15.0 	 \\ 
   36 &	    9 &	   35 &	  196 &	5.3 &	2.8 &	0.069 &	2.03 &	8.65 &	1.2 &	   77 &	14.3 	 \\ 
   37 &	  --- &	   37 &	 -107 &	5.1 &	2.5 &	0.068 &	1.90 &	6.88 &	1.9 &	   78 &	13.5 	 \\ 
   38 &	   10 &	   53 &	  -93 &	5.2 &	2.2 &	0.076 &	2.44 &	7.50 &	1.6 &	   75 &	 15.0 	 \\ 
   39 &	   11 &	   68 &	 -134 &	3.7 &	3.3 &	0.139 &	2.81 &	7.27 &	1.4 &	   78 &	13.5 	 \\ 
   40 &	   11 &	   88 &	 -160 &	3.2 &	2.3 &	0.069 &	1.23 &	7.28 &	0.8 &	   80 &	11.6 	 \\ 
   41 &	   11 &	  106 &	 -194 &	1.6 &	2.1 &	0.084 &	0.90 &	7.40 &	1.2 &	   80 &	11.6 	 \\ 
   42 &	   12 &	  108 &	  779 &	2.1 &	2.5 &	0.100 &	1.25 &	7.47 &	0.8 &	   81 &	11.8 	 \\ 
   43 &	  --- &	  138 &	  611 &	3.7 &	4.1 &	0.049 &	0.60 &	7.03 &	1.3 &	   82 &	11.9 	 \\ 
   44 &	   11 &	  142 &	 -238 &	0.9 &	1.6 &	0.075 &	0.41 &	7.52 &	0.9 &	   83 &	11.2 	 \\ 
   45 &	  --- &	  142 &	  493 &	1.2 &	1.8 &	0.062 &	0.37 &	7.38 &	0.9 &	   84 &	13.3 	 \\ 
   46 &	   12 &	  168 &	  630 &	1.6 &	2.0 &	0.098 &	0.98 &	7.19 &	0.4 &	  --- &	--- 	 \\ 
   47 &	   13 &	  181 &	 1263 &	1.3 &	2.3 &	0.075 &	0.54 &	7.15 &	0.7 &	   86 &	9.8 	 \\ 
   48 &	   12 &	  192 &	 1097 &	4.1 &	3.7 &	0.061 &	1.20 &	7.44 &	0.7 &	   87 &	10.4 	 \\ 
   49 &	   12 &	  221 &	  786 &	2.0 &	2.2 &	0.072 &	0.77 &	7.36 &	0.9 &	   88 &	10.3 	 \\ 
   50 &	   14 &	  311 &	  867 &	0.5 &	1.3 &	0.100 &	0.33 &	7.79 &	0.9 &	  --- &	--- 	 \\ 
\hline 
\end{tabular}  
\tablefoot{Column: (1) Core ID; (2) Fiber ID; (3) \& (4) central position in radio offsets respect to the map center, ($\Delta \alpha , \Delta \delta$); (5) 
peak intensity, I$_{peak}$ Intensity contrast respect to the surroundings, I$_{peak}/\mathrm{I}_0\ge$~30\%; 
 (7) equivalent FWHM=$\sqrt{FWHM_x^2+FWHM_y^2}$; (8) Mass within FWHM, M; (9) central velocity, V$_{lsr}$; (10) non-thermal velocity dispersion, $\sigma_{NT}/c_s$; (11) core ID in \citet{ROS08} catalog; (12) gas kinetic temperature, T$_K$ (when available). Cores are ordered by increasing $\Delta \alpha$.} 
\end{table*}

% fiber fragmentation
In addition to their quiescent kinematics, we notice that each of the observed fibers presents a high degree of fragmentation in the form of multiple condensations along the main axis. 
In cases like those well isolated fibers 1 or 12, this behaviour is already observed in the Herschel continuum maps
 (e.g., see Fig.~\ref{fig:NGC1333_PPV}c).
However, the identification of these condensations is limited in compact and crowded regions presenting line-of-sight superpositions.

\begin{figure*}[h]
   \centering
   \includegraphics[width=0.6\textwidth]{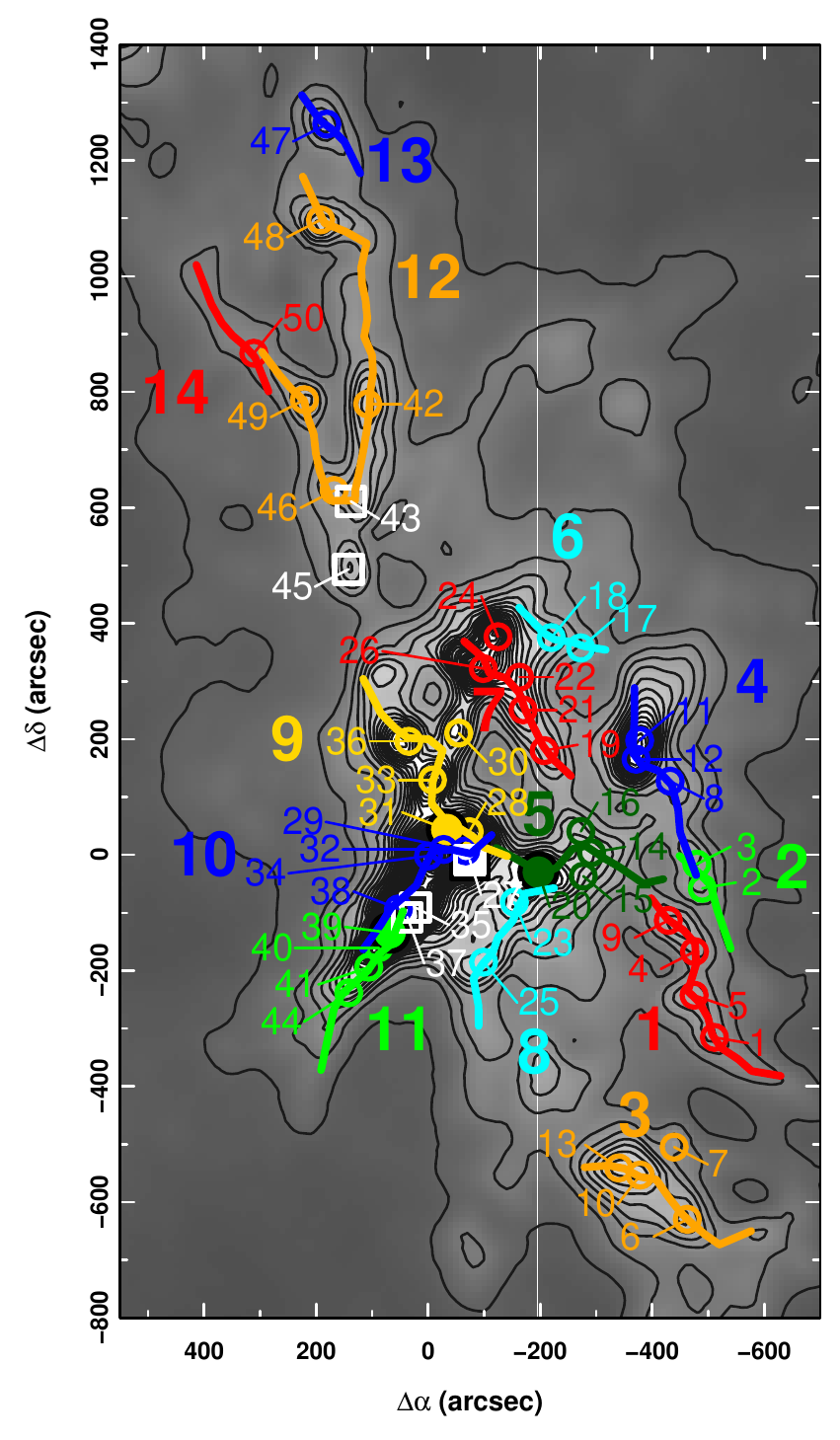}
      \caption{Core candidates identified in N$_2$H$^{+}$ in NGC1333 (labels 1-50; see Table~\ref{table:NGC1333_coreprop}). The position of the 44 cores associated to fibers are indicated by open circles and are color-coded similar to Fig.~\ref{fig:NGC1333_PPV}c. 
      For reference, the fiber numbers are labeled using a larger font.
      In addition, the 6 cores found in isolation are marked by white squares. Those cores with masses M~$\ge~2.5~M_\sun$ are highlighted with solid symbols. 
      }
             \label{fig:NGC1333_cores}
   \end{figure*}

We have investigated the internal fragmentation within the NGC1333 fibers using the reconstructed emission of these objects provided by our FIVE analysis. For each of the 14 fibers recovered, we have manually looked for  local maxima in their (recovered) total N$_2$H$^{+}$ emission maps. We identified as individual fragments all these local maxima with at least a 30~\% increase in emission with respect to the immediate surroundings, assumed as the mean intensity I$_0$ at a distance of 60 arcsec of these peaks, as expected for a centrally condensed, collapsing core entering in the non-linear regime \citep{INU97,HEI16}. When found, the characteristics of these local overdensities (central position, FWHM, mass, ...) are extracted using a 2D gaussian fit of the total integrated emission at these positions after subtracting an emission floor given by I$_0$. To complete our census, we also carry out this extraction process in those regions with detected N$_2$H$^{+}$ emission not associated to fibers. This simple approach allows us to easily resolve the gas emission in those regions with multiple velocity components, otherwise blended in the continuum maps. 
Different tests prove the robustness of this method in identifying the central position and peak intensity of the local maxima in our reconstructed N$_2$H$^{+}$ maps. Nevertheless, we notice a dependency of other fitting parameters (e.g., FWHM, total integrated intensity...) on the background subtraction and the fitting area. The use and interpretation of these last results should be taken with caution.

Using the above reconstruction technique, we identify a total population of 50 fragments detected in our N$_2$H$^{+}$ maps in NGC1333. Their main characteristics are listed in Table~\ref{table:NGC1333_coreprop}, including their position, intensity, mean FWHM, kinematic properties, and mass (cols. 3-10). Their positions are also indicated in Fig.~\ref{fig:NGC1333_cores}.
The 35 most prominent fragments correspond to those core positions identified in NH$_3$ by \citet{ROS08} (col. 9) with N$_2$H$^{+}$ counterparts within a radius of $\le$~50 arcsec and a velocity difference of $\le$~0.2~km~s$^{-1}$. Five of these cores are separated in two individual fragments in our sample, corresponding to those positions presenting a clear line multiplicity in NH$_3$ (e.g., core 80 in Rosolowski's catalogue). In addition, our analysis in velocity reveals another 11 fragments only detected in N$_2$H$^{+}$, hereafter referred to as core candidates. 

% Cores in fibers
The stable and quiescent properties of the pre-existing fibers detected in NGC1333 (see Sect.~\ref{sec:fertile_fibers}) seems to favour the conditions for their individual fragmentation.
Among those cores identified in our observations, 45 of them (90~\%) belong to points associated to fibers, while only 5 (10~\%) are found in isolation or associated to points not recovered by FIVE in such structures (see col.~2 in Table~\ref{table:NGC1333_coreprop}). Among them, we obtain an average number of 3-4 cores per fiber, with 12 out of the 14 (86\%) of these objects harbouring at least 2 or more cores. 

\begin{figure}
	\centering
	\includegraphics[width=0.9\linewidth]{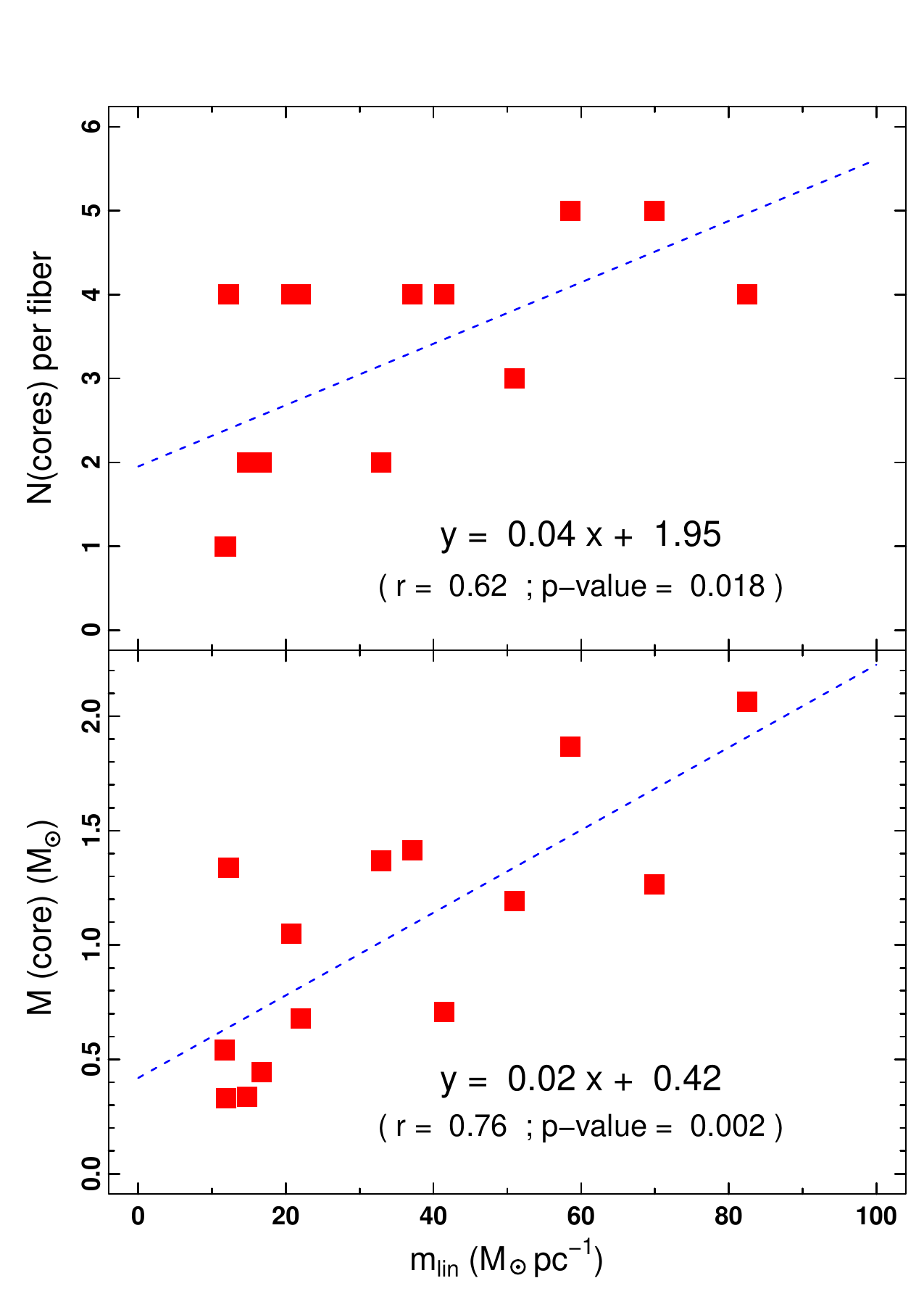}
	\caption{Fragmentation properties of the NGC1333 fibers (red solid points): Number of fragments (upper panel)  and  average core mass (lower panel) as a function of the fiber mass-per-unit-length (m$_{lin}$). The results of the linear fit of the data (blue dashed line), as well as their Pearson correlation coefficient (r) and the probability of the two distributions to be drawn from a random sample (p-value), are indicated in the lower right corner of each plot.}
	\label{fig:NGC1333_fibers_vs_cores}
\end{figure}

Statistically speaking, we observe a strong correlation between the properties of the cores and their parental fibers (see Fig.~\ref{fig:NGC1333_fibers_vs_cores}). 
We find a linear relationship between the mass per-unit-length of the fibers (m$_{lin}$) and the number of embedded cores in these objects (upper panel). Also, we observe a direct correlation between this same m$_{lin}$ and the average mass of the cores formed per fiber (lower panel).

In addition to these correlations, the regular core spacing found in several  fertile fibers (e.g., fibers 1 or 12) suggests that gravity is the main driver of their fragmentation. 
For cores belonging to the same fiber, we find an average core spacing of 0.08~pc, with values ranging between 0.05 and 0.20~pc (see Table~\ref{table:NGC1333_fiberprop}, last column). These results are consistent with the expected critical fragmentation scale for a filament in hydrostatic equilibrium $\lambda_{crit}=3.94\sqrt{\frac{2 c_s^2}{\pi G n(H_2)}}$ \citep{STO63} with  $\lambda_{crit}\sim$~0.1~pc for densities n(H$_2$)=~5~$\times$~10$^5$~cm$^{-3}$.

Originally proposed to explain the origin of cores in low-mass clouds \citep{HAC11,TAF15}, a core formation mechanism via  gravitational fragmentation of individual fibers might also be extended to the formation of cores in more massive and clustered environments. 
This fiber-fragmentation scenario differs from the classical hierarchical Jeans fragmentation by acting only in the individual fibers and not at cloud scales\footnote{Estimations of the fragmentation scale based on the average distances between cores/stars without considering their association to their harbouring fibers are, therefore, meaningless in this fiber framework.}. 
The gravitational fragmentation process of the NGC1333 fibers appears to be controlled by the global properties of their parental fibers \citep{TAF15}, where the most massive fertile fibers efficiently form more massive cores and in higher numbers. 

% Spatial distribution
Due to their origin, the spatial distribution of cores and young protostars are directly related to the spatial distribution of fibers in NGC1333. We have investigated this property from the surface density of cores and its comparison with the position of fibers. As demonstrated in Fig.~\ref{fig:NGC1333_KDE2D}e, the highest concentrations of cores (black contours) are found towards those regions with a higher number of fibers (red segments).  Additionally, the position of the cores and fibers also coincides with the distribution of the youngest Class 0/I objects (black contours). In the light of the above, we conclude that the internal mass distribution of the NGC1333 proto-cluster is primarily determined by the position and orientation of its fiber substructure.

%\subsection{Stellar feedback and fiber destruction}

\section{Discussion}\label{sec:discussion}

\subsection{Clusters as complex fiber networks}

Our analysis of the internal gas kinematics traced in N$_2$H$^+$ in Sect.~\ref{sec:kinematics} shows how the dense gas within NGC1333 is highly structured forming an intricate network of 14 quiescent fibers. These objects are characterized by an internal velocity-coherence and have (trans-)sonic internal motions at scales of L$_{fib}\sim$~0.4~pc. In Sect.~\ref{subsec:kinematics:fibers}, we demonstrated that the fragmentation and collapse of these pre-existing structures control the current formation of stars in this proto-cluster. 
Independent results suggest a similar level of substructure in most of the intermediate-mass proto-clusters in the solar neighbourhood. \citet{MOT98} have identified several series of filamentary and elongated chains of cores within the \object{$\rho$~Oph} cluster at scale of L$_{fib}\sim$~0.2~pc at the continuum. When observed with dense gas tracers like N$_2$H$^+$, these chains of cores appear to be connected in velocity showing (tran-)sonic velocity dispersions  \citep{DIF04,AND07}. Twisting filamentary structures at scales of L$_{fib}\sim$~0.5~pc have been also reported within the \object{Serpens South} proto-cluster by \citet{KIR13}, coincident with regions showing complex N$_2$H$^+$ spectra detected in different single-dish data \citep{TAN13} . Using CARMA interferometric observations, \citet{FER14} have identified these small-scale filaments as fiber-like structures based on their velocity continuity and internal motions. 
Similarly, nearby clusters are typically found at the centre of filamentary hubs \citep{MYE09} forming complex networks of sub-filaments (aka fibers) in regions like \object{B59} \citep{ROM09}, \object{L1517} \citep{HAC11}, L1495 \citep{HAC13}, and \object{Serpens Main} \citep{LEE14,ROC15}.  
 
While some local differences are expected in more compact environments, increasing evidence suggest that this complex gas organization could also be extended to more massive regimes.
The NGC1333 fibers resemble the so-called molecular fingers detected in the \object{OMC-1} region \citep{MART90,ROD92}. Despite being embedded within the Orion Nebula Cluster, these molecular fingers exhibit a  
surprising level of coherence characterized by their low internal motions at scales of L$_{fib}\sim$~0.2-0.3~pc \citep{WIS98}. 
In addition, recent high angular resolution interferometric observations have indicated the presence of filamentary networks within several Infrared Dark Clouds and massive star-forming clumps \citep{PER13,ZHA15,HEN16}. 
The inhomogeneous analysis and distinct nature of these data prevent a direct inter-comparison of these fiber-like structures. 
Nevertheless, and based in the strong similarities with the results presented in this work, the presence of similar fiber networks appear to explain the internal structure of high-mass proto-clusters.

\subsection{Clustered vs. Isolated star formation}

In Sect.~\ref{subsec:kinematics:fibers}, we demonstrated how the NGC1333 fibers present roughly similar sizes and kinematic properties to those previously identified in regions like B213-L1495 in Taurus. As discussed in Sect.~\ref{sec:fertile_fibers}, the larger linear masses of the NGC1333 fibers are attributed to the choice of N$_2$H$^+$ as gas tracer leading to an observational bias towards objects presenting higher densities. Despite these selection effects, the range of masses and characteristics of the NGC1333 fibers coincide with those reported for the sub-group of star-forming, fertile fibers in B213-L1495 (see Fig.~\ref{fig:NGC1333_histos}c). 
The strong similarities between the two regions contrasts with the remarkably diverging star-formation activities in these two clouds. 
NGC1333 has formed a total of N(proto)=~45 protostars in the last Myr within a compact area of $\lesssim$~1~pc$^2$, while B213-L1495 was (only) able to form N(proto)$\sim$19 young stellar objects along $>$~10~pc$^2$ within the same period of time.
This difference rises a fundamental question for our description of the star formation process in both isolated (e.g., Taurus) and clustered (e.g., Perseus) star-forming regions: if both types of clouds are formed by the same type of fibers, how is the star formation process in the NGC1333 proto-cluster enhanced in comparison to the B213-L1495 cloud? 

In Table~\ref{table:comparison}, we summarize the main physical characteristics of both B213-L1495 and NGC1333 regions (cols. 1 \& 2). As deduced from their absolute ratios (col.~3), neither their total mass nor their total number of fibers (i.e., N(fibers)) explain the current star formation activities of these two regions. Similarly, the differences between the clouds cannot be solely attributed to a more effective fragmentation (i.e., N(cores/fertile fiber)) nor a higher gas efficiency (i.e., SFE(dense)).
Rather, we find an excellent agreement between the relative numbers of protostars N(proto) and cores N(cores) with the number fertile fibers N(fertile fibers) in these two regions, with differences of $<$~30~\%. These values translate into similar ratios for their corresponding surface densities (aka, $\Sigma$(protostars) vs. $\Sigma$(cores) vs $\Sigma$(fertile fibers)). Based on this comparison, we conclude that the higher surface density of protostars and cores in NGC1333 results from a higher surface density of fertile fibers in this clustered region respect to the isolated B213-L1495 cloud.

\begin{table} 
\label{table:comparison} 
\caption{Star formation in NGC1333 and B213-L1495} 
\centering 
\begin{tabular}{l c c c c c } 
%\hline 
\hline 
  & B213 & NGC1333 &	Ratio \\ 
  & (1) & (2) & (3) \\
\hline
Length (pc) $^\star$  & $\sim$ 10 & $\sim$ 2 & --- \\
Radius (pc) $^\star$  &  $\sim$ 0.5  & $\sim$ 0.4 & --- \\
Area=$L\times 2R$ (pc$^{2}$) & 10 & 0.8 & 0.08 \\
%\hline
Total mass (gas+stars; M$_\sun$) & $\sim$~700 & $\sim$~1300 & 1.8 \\
\hline
N(protostars)  & 16 & 45 & 2.8 \\
N(cores)       & 19 & 50 & 2.6 \\
N(fibers) & 35 & 15 & 0.4 \\
N(fertile fibers) & 7 & 15 & 2.1 \\
N(cores)/N(fertile fib.) & 2.7 & 3.7 & 1.4 \\
SFE(dense) (\%) & $\sim$ 7.5 & $\sim$ 10 & 1.3  \\
\hline
$\Sigma$ (proto) (pc$^{-2}$)$^\star$$^\star$ & 1.6 & 56.3 & 35.1 \\ 
$\Sigma$ (cores) (pc$^{-2}$)$^\star$$^\star$ & 1.9 & 62.5 & 32.9 \\
$\Sigma$ (fertile\ fibers) (pc$^{-2}$)$^\star$$^\star$ & 0.7 & 18.7 &  26.8 \\
\hline 
\end{tabular}  
\tablefoot{(1) values derived by \citet{HAC13}; (2) this work; (3) ratio between NGC1333 and B213-L1495. Remarks:
($\star$) Approximate cloud extension (aka length and radius) covered in the corresponding studies.
 ($\star$$\star$) Surface densities calculated assuming with $\Sigma (X)=N(X)/(2RL)$
} 
\end{table} 

Our direct comparison of the results in Taurus and Perseus suggests that the differences between the classically distinguished isolated and clustered star formation scenarios might be explained from the selection and spatial density of fertile fibers. If the main properties of these last objects remain independent of the environment, the number of fertile fibers and their surface density should linearly increase with the number and density of protostars in a star-forming region, with $\Sigma$ (proto)~$\propto$~$\Sigma$ (fertile\ fibers), in agreement to our results in NGC1333.
If confirmed, the formation of stars in clusters would then naturally arise as an scaled-up version of those formed in isolated star forming clouds. Observations in regions of increasing mass, complexity, and stellar activity (e.g., Orion) are needed to confirm this hypothesis (Hacar et al in prep.).

\subsection{Fiber-to-fiber collisions and the origin of massive cores}

Despite all the caveats in our mass estimates, an apparent correlation seems to emerge between the internal distribution of fibers and the position of the most massive cores and stars in NGC1333.
Highlighted by solid symbols in Fig.~\ref{fig:NGC1333_cores}, the four most massive cores in this cloud, namely, cores 20, 27, 31, and 39, all with M~$>$~2.5~M$_\sun$, are found at junction of multiple fertile fibers and coincident with the positions of the IRAS 2, 3, and 4 sources. Less clear, although still appealing,
the location of the two B-type BD+30$º$549 and IRAS~8 stars seems to correspond to the centre of a fan-like arrangement of fibers. 
Given these results, could the interaction between fibers be responsible of the formation of massive cores and stars? 

If the spatial density of fibers increases in more crowded and clustered environment, the size L$_{fib}$ and radius R$_{fib}$ of these objects determine the maximum level of compactness of the fiber networks. For a cylindrical volume of $\pi$R$_{0}^2$~L$_{fib}$, and assuming a perfect bundle of parallel fibers, it is easy to prove that the maximum number of fibers that can be packed together (without interaction) is limited to N$_{max}\le (R_0/R_{fib})^2$ fibers. Higher spatial (3D) densities of fibers would lead to the direct collision of several of these gaseous objects. %A similar core-to-core collision scenario has been proposed for the formation of massive stars in dense clusters. 
With a larger cross-section than cores, the elongated nature of the fibers in combination with a more random orientation would also promote their direct interaction, rapidly decreasing of N$_{max}$. In compact and clustered environments with R$_0\lesssim$~L$_{fil}$ and multiple intertwined fibers like NGC1333, fiber-to-fiber collisions appear to be practically unavoidable.

In the prevalent filamentary paradigm of star formation, a vast theoretical framework describes the core formation mechanisms via filament fragmentation \citep[e.g.,][]{LAR85,NAG87,INU97,FIE00,HEI16}.
According to these models, the mass of the cores is determined by the evolution of Jeans-unstable density perturbations within individual filaments. In this framework, a cut-off in the expected core mass distribution is, however, imposed by the linear mass of their parental filaments, originally close to equilibrium \citep{INU01}.  
Collisions between fibers offer an interesting alternative to this fragmentation scenario in complex fiber networks. Mergings of two (or more) fibers would generate large overdensities leading to the formation of massive and highly unstable cores. The frequency of these mergings would increase with the network complexity and the surface density of fibers. Intuitively, and in comparison to isolated fibers fragmenting into Jeans-like cores, denser clusters with higher number of fibers would naturally develop larger numbers and more massive, super-Jeans cores.

\section{Conclusions}

We have investigated the dense gas properties and internal substructure of one the nearest embedded, gas-rich proto-clusters in the solar neighbourhood, the NGC1333 region in Perseus. Our observational study combines a series of new large-scale, density selective IRAM~30m N$_2$H$^+$ (1-0) and Effelsberg~100m NH$_3$ (1,1) and (2,2) molecular line observations along with previously published Herschel-Planck dust-continuum data. The results we obtained are as follow:

\begin{enumerate}

\item For the entire NGC1333 ridge, we estimate a total mass of gas of $\sim$~1700 M$_\sun$ based on the Herschel continuum maps. This gas component dominates the total mass load of this embedded cluster at all radii ($\gtrsim 80$\%) compared to its stellar contribution (< 20\%). 

\item  Based on our  N$_2$H$^+$ and NH$_3$ observations, we detect a total mass of
$\sim$~250 M$_\sun$ of dense gas at densities between n(H$_2$)~$\sim 5\times 10^4 -  10^6$~cm$^{-3}$. This dense component corresponds to the coldest (T$_{eff}\lesssim$~15~K) and highest column density material (N(H$_2$)~$\gtrsim$~10$^{22}$~cm$^{-2}$) within the NGC1333 region.  We report an increase of both the NH$_3$-derived gas kinetic and Herschel dust effective temperatures ($>$~15~K) in the surroundings of the most active IRAS sources. Despite these local heating effects, most of the dense gas remains unaffected by the stellar feedback, and traces the pristine gas structure of this young proto-cluster.

\item An inspection of the N$_2$H$^+$ line profiles reveals a complex internal gas kinematics characterized by a high multiplicity and line variability. Using a improved version (v.1.1) of the FIVE algorithm \citep{HAC13}, we identify a total of 14 velocity-coherent fibers within the central NGC1333 clump. The mass recovered in fibers corresponds to $\sim$~90\% of the total mass of dense gas in this cloud detected in spectra with S/N~$\geq$~3.

\item Statistically speaking, the NGC1333 fibers present total lengths of $\sim$~0.4~pc, transonic internal velocity dispersions, and mass-per-unit-length between m$_{lin}\sim$~12-85 M$_\sun$~pc$^{-1}$. Also, the fibers exhibit a high degree of internal fragmentation with typically 3-4 cores per fiber. Their average properties mimic those of the so-called fertile fibers previously identified in low-mass fibers identified in B213-L1495 using similar extraction and analysis techniques. Based on its internal gas properties and substructure, we define the NGC1333 proto-cluster as a bundle of fertile fibers.

%\item Within the NGC1333 proto-cluster, most of the cores seem to be originated from the gravitational fragmentation of fibers. Among the 50 core candidates identified in this region, 45 (90~\%) belong to points associated to this primordial gas structure. We find direct correlations between the mass and number of embedded cores with the mass-per-unit-length of their parental fibers. Moreover, and within each of these objects, cores are regularly spaced with distances consistent with the critical fragmentation scale for a gaseous cylinder at densities n(H$_2$)~$\sim 5\times 10^5$~cm$^{-3}$. 

\item Our observations reveal both isolated star-forming clouds \citep[e.g., B213-L1495,][]{HAC13} and intermediate-mass clusters (e.g., NGC1333, this work) as fiber networks of different complexity.
In both types of clouds,  most of the dense cores (>90~\%) are originated within fibers, likely via gravitational fragmentation. 

\item For the first time, our calculations demonstrate that the current number and surface density of dense cores and young protostars linearly scales with their corresponding observed number and surface density of fertile fibers, respectively. These results suggest that the classical distinction made between isolated (e.g., Taurus) and clustered (e.g., Perseus) star-forming clouds might originate from their initial density of fertile fibers. 

\end{enumerate}

\begin{acknowledgements}
A.H. thanks Alex Kraus and Benjamin Winkel their support during the observations at the Effelsberg 100m telescope.
This work has benefited from research funding from the European Community's Seventh Framework Programme.
This paper has benefit from the useful discussions with Pavel Kroupa and Andreas Burkert.
This work is part of the research programme VENI with project number 639.041.644, which is financed by the Netherlands Organisation for Scientific Research (NWO).
MT and AH thank the Spanish MINECO for support under grant
AYA2016-79006-P.

\end{acknowledgements}

%-------------------------------------------------------------------

\end{document}